\documentclass[preprint,pteplogo]{ptephy_v2}

\preprintnumber{XXXX-XXXX} 
\usepackage{hyperref}

\usepackage{bm}
\usepackage{graphicx}
\usepackage{ascmac}
\usepackage{amsmath,amssymb}
\usepackage{cancel}
\usepackage{braket}
\usepackage{lineno}
\renewcommand{\Re}{\operatorname{Re}}
\renewcommand{\Im}{\operatorname{Im}}
\usepackage{subfigure}

\begin{document}

\title{Compositeness of near-threshold $s$-wave resonances}


\author{Tomona Kinugawa}
\affil{Few-body Systems in Physics Laboratory, RIKEN Nishina Center, Wako 351-0198, Japan \email{tomona.kinugawa@riken.jp}\\}

\author{Tetsuo Hyodo}
\affil{Research Center for Nuclear Physics, The University of Osaka, Ibaraki, Osaka 567-0047, Japan}




\begin{abstract}
The near-threshold clustering phenomenon is well understood by the low-energy universality, for shallow bound states below the threshold. Nevertheless, the characteristics of resonances slightly above the threshold still lack thorough elucidation. We introduce a novel probabilistic interpretation scheme for complex compositeness of resonances, in which resonances with large decay widths fall outside the domain where their internal structure can be interpreted probabilistically. Employing this scheme to analyze resonances via the effective range expansion, we demonstrate that narrow interpretable near-threshold resonances have a small composite fraction, in sharp contrast to shallow bound states below the threshold.
\end{abstract}

\subjectindex{xxxx, xxx}

\maketitle

\section{Introduction}

In nature, the hierarchical structure is formed by the particles in different energy scales, such as quarks, hadrons, nuclei, and atoms. In each hierarchy, the clustering phenomenon is known to emerge universally, leading to the formation of subunits that act as effective degrees of freedom~\cite{Nakamura:2025ivk}. For example, while the fundamental degree of freedom in nuclear hierarchy is the nucleon, the ${}^{12}$C Hoyle state consists of three $\alpha$ particles (${}^{4}$He nuclei) as the subunits~\cite{Hoyle:1954zz,PTPS52.89}. In hadron physics, in addition to the ordinary hadrons composed of quarks as the fundamental particles, it is considered that loosely bound systems of hadrons, the hadronic molecules, manifest as alternative internal structures~\cite{Guo:2017jvc}. Such exotic internal structure of hadronic molecules has recently attracted a lot of attention from researchers~\cite{Hosaka:2016pey,Brambilla:2019esw}.

To quantitatively characterize the cluster structure of hadrons, the notion of compositeness has been introduced and subsequently developed~\cite{Weinberg:1965zz,Baru:2003qq,Hyodo:2011qc,Aceti:2012dd,Xiao:2012vv,Aceti:2014ala,Hyodo:2013iga,Chen:2013upa,Hyodo:2013nka,Sekihara:2013sma,Aceti:2014wka,Nagahiro:2014mba,Hyodo:2014bda,Sekihara:2014qxa,Sekihara:2014kya,MartinezTorres:2014kpc,Navarra:2015iea,Garcia-Recio:2015jsa,Meissner:2015mza,Guo:2015daa,Kamiya:2015aea,Sekihara:2015gvw,Guo:2016wpy,Lu:2016gev,Kang:2016ezb,Kamiya:2016oao,Kang:2016jxw,Tsuchida:2017gpb,Oller:2017alp,Matuschek:2020gqe,Esposito:2021vhu,Li:2021cue,Du:2021zzh,Song:2022yvz,Albaladejo:2022sux,Mikhasenko:2022rrl,Kinugawa:2022fzn,vanKolck:2022lqz,Kinugawa:2023fbf,Dai:2023cyo,Song:2023pdq,Yin:2023wls,Kinugawa:2024crb,Terashima:2025tbz,Kinugawa:2026fob}. The compositeness of a stable bound state refers to the probability of finding the composite component in the wave function. This concept has been utilized across the hierarchical structure to analyze the deuteron, hypertriton and ${}^{4}$He dimer~\cite{Weinberg:1965zz,Li:2021cue,Song:2022yvz,Kinugawa:2022fzn,Yin:2023wls,Kinugawa:2024crb}. The compositeness is also expected to be useful in quantitatively extracting the molecular component from the wave function of exotic hadrons which are the superposition of various components. However, the majority of exotic hadrons decay with finite lifetime. This instability leads to the complex-valued compositeness~\cite{Hyodo:2013nka,Kinugawa:2024crb}, posing challenges in interpretation. This is the first issue which we want to discuss in this paper.

Another remarkable fact about exotic hadrons is their appearance near the two-body threshold energy. Near-threshold phenomena follow the low-energy universality, regardless of the details of the system, particularly when the magnitude of the scattering length is sufficiently larger than other characteristic length scales of the system~\cite{Braaten:2004rn,Naidon:2016dpf}. Universality dictates that the near-threshold $s$-wave bound states below the threshold are typically composite dominant, indicating the clustering phenomena~\cite{Hyodo:2014bda,Hanhart:2014ssa,Sazdjian:2022kaf,Lebed:2022vks,Hanhart:2022qxq,Kinugawa:2023fbf}. In contrast, although the nature of near-threshold $s$-wave resonances above the threshold has been discussed~\cite{Matuschek:2020gqe,Habashi:2020qgw,Habashi:2020ofb}, it has not yet been fully elucidated. However, recent analysis of the $K^{-}\Lambda$ correlation function data for $\Xi(1620)$~\cite{Sarti:2023wlg} suggests the existence of such a resonance. In nuclear physics, the possible existence of ${}^{21}$C as an $s$-wave resonance of the ${}^{20}$C$n$ system has also been suggested~\cite{Platter:2025pmk}. Furthermore, $s$-wave resonances can also be realized in cold-atom physics via the Feshbach resonance induced by the external magnetic field~\cite{Kohler:2006zz}. Therefore, the second issue addressed in this paper is to study the nature of near-threshold $s$-wave resonances above the threshold.

Our objective in this paper is to offer a plausible interpretation of the complex compositeness and to provide insights into the nature of near-threshold $s$-wave resonances. As a prerequisite for investigating the internal structure of resonances in terms of the compositeness, we first examine in Sec.~\ref{sec:eigenstates} whether the wave function of an eigenstate admits a meaningful interpretation in terms of its internal structure. In Sec.~\ref{sec:ERE}, near-threshold resonances are then examined within the framework of the effective range expansion, taking into account the low-energy universality. Subsequently, we propose a probabilistic interpretation of the complex compositeness in Sec.~\ref{sec:interpretation}. As the main result of this work, Sec.~\ref{sec:structure} quantitatively demonstrates that interpretable near-threshold $s$-wave resonances are not composite dominant, in contrast to near-threshold bound states. This result is compared with those obtained using other prescriptions in previous studies in Sec.~\ref{sec:comparison}. The proposed interpretation scheme is applied to hadronic resonances in Sec.~\ref{sec:apply}. Finally, Sec.~\ref{sec:sum} summarizes this work.

\section{Eigenstates and wavefunctions}
\label{sec:eigenstates}

Before discussing the structure of near-threshold resonances in terms of the compositeness, this section first clarifies how the internal structure of an eigenstate can be related to the properties of its wave function. Since unstable states have properties different from those of stable states, it is nontrivial whether their internal structure can be discussed in the same way as that of bound states. In fact, it is not even clear whether unstable states possess an internal structure that can be characterized in terms of the compositeness. Here, we adopt the viewpoint that the internal structure of a state is reflected in the properties of its wavefunction, and examine the difference between unstable and stable states from this perspective. This allows us to distinguish states that admit a meaningful internal-structure interpretation from those that do not admit such an interpretation.

The eigenstates of the Hamiltonian are classified according to their complex eigenmomenta~\cite{Taylor}. The eigenmomentum of a stable bound state is purely imaginary with a positive imaginary part. On the other hand, an unstable resonance has a complex eigenmomentum with a positive real part and a negative imaginary part. In addition to these states, a state with a purely imaginary eigenmomentum whose imaginary part is negative is called a virtual state.

Here, in order to examine the wavefunctions of the respective states explicitly, we consider the $s$-wave eigenstates in a three-dimensional spherically symmetric square-well potential. A bound state exists when the interaction is sufficiently attractive. As the attraction is weakened, the bound state becomes shallower and eventually turns into a virtual state. On the other hand, resonances exist in an infinite number, irrespective of the sign of the interaction~\cite{NP11.499}. In the following, we discuss the spatial localization of the probability density based on the $s$-wave radial wavefunction.

\begin{figure}[tb]
\centering
\includegraphics[width=0.45\textwidth]{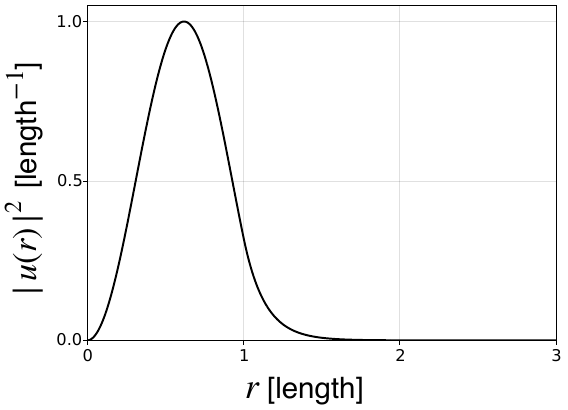}
\includegraphics[width=0.45\textwidth]{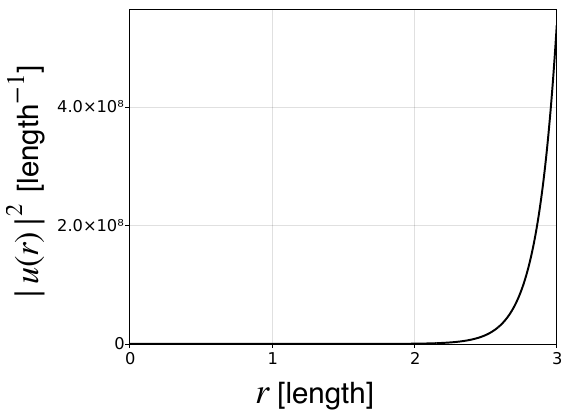}
\caption{Wavefunctions of a bound state (left) and virtual state (right).}
\label{fig:bound-virtual}
\end{figure}

In the case of a bound state, the wavefunction is localized around the origin (Fig.~\ref{fig:bound-virtual}, left). The localization of the wavefunction near the origin indicates that its probability density is spatially concentrated within a finite interaction region. Therefore, we consider that it is possible to discuss the internal structure of bound states, because the wave function describes a spatially localized object. From this perspective, the internal structure of virtual states and resonances can be discussed in the same way as that of bound states, only when their wavefunctions are similarly localized.

In contrast to bound states, however, the wavefunctions of virtual states and resonances are known to diverge at large distances. In fact, as shown in Fig.~\ref{fig:bound-virtual}, right, the wavefunction of the virtual state diverges. Furthermore, that wavefunction does not exhibit localization in the interaction region. In other words, bound states and virtual states possess qualitatively different wavefunctions, and thus their internal structure cannot be discussed on an equal footing. 

\begin{figure}[t]
\centering
\includegraphics[width=0.45\textwidth]{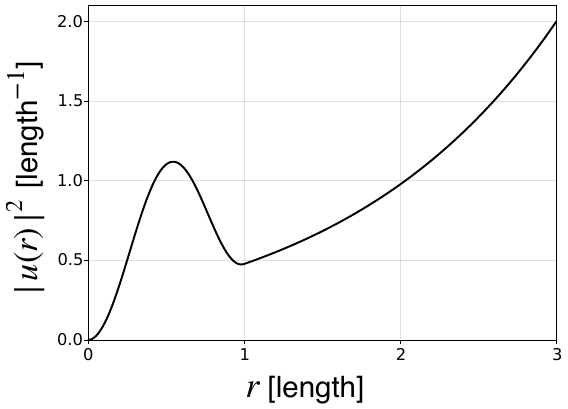}
\includegraphics[width=0.45\textwidth]{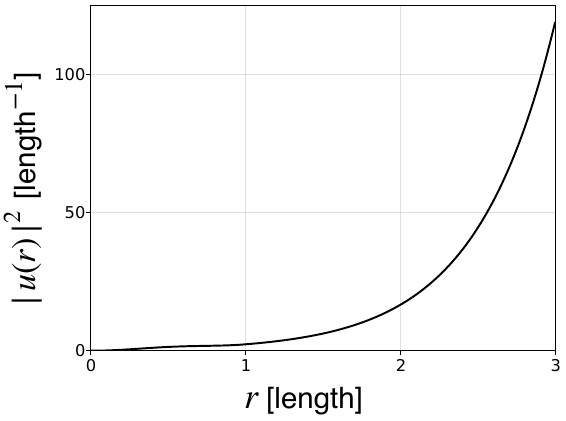}
\caption{Wavefunctions of a narrow resonance (left) and broad resonance (right).}
\label{fig:resonances}
\end{figure}

We then consider resonances. For a resonance pole with the complex energy $E = E_{R}-i\Gamma/2$, the decay width is given by $\Gamma=-2{\rm Im}\, E$. In the following, we characterize a resonance as narrow or broad in terms of the ratio of its decay width to the excitation energy $\Gamma/E_R$. We show the wavefunctions of a narrow resonance with $\Gamma/E_{R} \sim 0.27$ (Figure~\ref{fig:resonances}, left) and a broad resonance with $\Gamma/E_{R} \sim 1.27$ (Figure~\ref{fig:resonances}, right). As seen in the figure, both wavefunctions diverge at large distances. Nevertheless, the narrow resonance exhibits localization near the origin. We therefore conclude that only narrow resonances admit a meaningful internal-structure interpretation. This is in line with the fact that, in the limit $\Gamma\to 0$, a resonance approaches a bound state in character.

Based on these considerations, we regard bound states and narrow resonances as isolated states whose internal structure can be reasonably interpreted. On the other hand, we exclude virtual states and broad resonances from this category, because their wavefunctions do not localize and thus do not represent spatially localized objects whose internal structure can be meaningfully discussed. This does not mean that the study of virtual states or broad resonances themselves is meaningless. It is legitimate that poles in the scattering amplitude, such as those corresponding to the $\sigma$ meson in the $\pi\pi$ scattering or the virtual state in the ${}^1S_0$ channel of the $NN$ scattering, can exist and may even affect physical observables. Since this study focuses on the interpretation of internal structure, we will propose an interpretation scheme in which only bound states and narrow resonances are considered interpretable. 

\section{Nature of near-threshold resonances}
\label{sec:ERE}

To discuss the internal structure of near-threshold resonances, we employ the effective range expansion (ERE), which provides a model-independent description of low-energy scattering. In this section, we summarize the properties of near-threshold resonances in the ERE~\cite{Hyodo:2013iga}. Although hadron resonances generally appear in coupled-channel systems, we restrict the present discussion to a single-channel system in order to elucidate the essential properties of near-threshold $s$-wave resonances in the simplest setting. An extension incorporating the effects of additional open channels can in principle be formulated straightforwardly. The single-channel $s$-wave scattering amplitude $f(k)$ with a small momentum $k$ is expressed only by the scattering length $a_{0}$ and effective range $r_{e}$:
\begin{align}
f(k)&=\left[-\frac{1}{a_{0}}+\frac{r_{e}}{2}k^{2}+\mathcal{O}(k^{4})-ik\right]^{-1}.
\label{eq:f-ERE}
\end{align}
Discrete eigenstates of the Hamiltonian of the system are expressed by the poles of $f(k)$ with the analytically continued momentum $k$ in the complex plane~\cite{Kukulin,Moiseyev,Hyodo:2020czb}. By solving the pole condition $1/f(k)=0$, we obtain two poles $k^{\pm}$ as
\begin{align}
k^{\pm}&=\frac{i}{r_{e}}\pm\frac{1}{r_{e}}\sqrt{\frac{2r_{e}}{a_{0}}-1+i0^{+}}.
\label{eq:k-pm}
\end{align}
Regardless of the signs of $a_{0}$ and $r_{e}$, the position of $k^{-}$ is always closer to the physical scattering region (Re $k\geq 0$ and Im $k=0$). This physically important $k^{-}$ is called the primary pole. Let us present useful expressions of the scattering length $a_{0}$ and effective range $r_{e}$ in terms of the eigenmomenta. Conversely, from Eq.~\eqref{eq:k-pm}, the scattering length and effective range are expressed in terms of $k^{\pm}$ as~\cite{Hyodo:2013iga}
\begin{align}
a_{0}&=-\frac{k^{+}+k^{-}}{ik^{+}k^{-}},\quad 
r_{e}=\frac{2i}{k^{+}+k^{-}} .
\label{eq:r-ERE}
\end{align}

In this paper, we consider the system with a negative effective range to realize the resonance solution~\cite{Hyodo:2013iga}. Here we note that the $k^{2}$ term in Eq.~\eqref{eq:f-ERE} cannot be neglected to describe the resonance pole, because the resonance always appears with the anti-resonance, and the pole condition should be at least quadratic in $k$. In this sense, the contribution from $r_{e}$ is essential for near-threshold resonances. 

Because the resonance pole appears in the $-\pi/4\leq \theta_{k}< 0$ region with $k^{-}=|k^{-}|e^{i\theta_{k}}$, we should have Re~$k^{-}\geq|$Im~$k^{-}|$. In this case, Eq.~\eqref{eq:k-pm} gives the following relation:
\begin{align}
\frac{1}{|r_{e}|}\sqrt{\frac{2r_{e}}{a_{0}}-1}& \geq \frac{1}{|r_{e}|},\quad 
\Rightarrow \quad \frac{r_{e}}{a_{0}} \geq 1 .
\end{align}
From this inequality and the condition of negative $r_{e}$, $a_{0}$ is found to be also negative. We then find that $|r_{e}|$ is always larger than $|a_{0}|$ for resonances, no matter how large $|a_{0}|$ is:
\begin{align}
r_{e} \leq a_{0} < 0,\quad \Rightarrow \quad |a_{0}|\leq |r_{e}|.
\label{eq:a-r-relation}
\end{align}
In other words, the scattering length is not the unique scale for the near-threshold resonance because of the non-negligible effective range, and the low-energy universality does not hold. In fact, we find that not only $|a_{0}|$ but also $|r_{e}|$ diverge in the $|k^{-}| \to 0$ limit by keeping the relation~\eqref{eq:a-r-relation}. To see this, we express $a_{0}$ and $r_{e}$ by $|k^{-}|$ and $\theta_{k}$ ($k^{-} = |k^{-}|e^{i\theta_{k}}$) as follows:
\begin{align}
a_{0}&=-\frac{2\sin\theta_{k}}{|k^{-}|},\quad
r_{e}=-\frac{1}{|k^{-}|\sin\theta_{k}}.
\label{eq:a-re-k}
\end{align}
These equations show that both $|a_{0}|$ and $|r_{e}|$ diverge in the $|k^{-}| \to 0$ limit in the case of resonance $-\pi/4\leq \theta_{k}<0$. From this viewpoint, we find that near-threshold resonances are not necessarily composite-dominant in contrast to near-threshold bound states.

When a resonance--anti-resonance pair appears, complex $k^{\pm}$ satisfy the relation $k^{+}=-k^{-*}$, with which we can rewrite $a_{0}$ and $r_{e}$ only with the eigenmomentum of the resonance $k^{-}$:
\begin{align}
a_{0}&=\frac{2\Im k^{-}}{|k^{-}|^{2}}=\frac{2\Im k^{-}}{(\Re k^{-})^{2} + (\Im k^{-})^{2}},\quad 
r_{e}=\frac{1}{\Im k^{-}}.
\label{eq:r-ERE-km}
\end{align}
When Re $k^{-}=-$Im $k^{-}$, $a_{0} = r_{e}$ follows from these equations, corresponding to the equality in Eq.~\eqref{eq:a-r-relation}. This occurs when the real part of the eigenenergy is zero. 

The compositeness of a stable bound state is introduced by expressing its wave function as a superposition of $\ket{{\rm composite}}$ and $\ket{{\rm elementary}}$. The probability of finding $\ket{{\rm composite}}$ ($\ket{{\rm elementary}}$) is defined as the compositeness $X$ (elementarity $Z=1-X$)~\cite{Weinberg:1965zz,Hyodo:2013nka,Kamiya:2015aea,Kamiya:2016oao,Kinugawa:2024crb}. The compositeness of near-threshold states is obtained from the weak-binding relation written in terms of $a_{0}$ and $r_{e}$~\cite{Weinberg:1965zz,Hyodo:2013nka,Hyodo:2013iga,Oller:2017alp,Matuschek:2020gqe,vanKolck:2022lqz}:
\begin{align}
X&=\sqrt{\frac{1}{1-\frac{2 r_{e}}{a_{0}}}} .
\label{eq:wbr}
\end{align}
For the recent discussion on the finite range correction to Eq.~\eqref{eq:wbr}, see Refs.~\cite{Li:2021cue,Song:2022yvz,Albaladejo:2022sux,Kinugawa:2022fzn,Kinugawa:2023fbf,Yin:2023wls}. 

By substituting $a_{0}$ and $r_{e}$ in Eq.~\eqref{eq:a-re-k} into Eq.~\eqref{eq:wbr}, we obtain the expression of the compositeness $X$ in terms of the pole positions~\cite{Hyodo:2013iga}. With the relation $k^{+}=-k^{-*}$ for resonances, the compositeness of resonances can be expressed only by the argument of the eigenmomentum $\theta_{k}$ or by that of the eigenenergy $\theta_{E}$:
\begin{align}
X&=-i\tan \theta_{k}=-i\tan\left(\frac{1}{2}\theta_{E}\right).
\label{eq:X-ERE-theta}
\end{align}
This also shows that the compositeness of near-threshold resonances is pure imaginary~\cite{Hyodo:2013iga}. 
 
\section{Interpretation scheme for complex compositeness}
\label{sec:interpretation}

In general, the compositeness of resonances is defined as a complex value~\cite{Kinugawa:2024crb}. In fact, the compositeness of resonances in the ERE is pure imaginary as shown in Eq.~\eqref{eq:X-ERE-theta}. Thus, to consider the internal structure of resonances, we need to consider its probabilistic interpretation. In the previous works~\cite{Aceti:2012dd,Xiao:2012vv,Aceti:2014ala,Hyodo:2013iga,Hyodo:2013nka,Sekihara:2013sma,Kamiya:2015aea,Sekihara:2015gvw,Kamiya:2016oao,Matuschek:2020gqe}, some prescriptions were proposed by introducing a fraction associated with the composite component in the resonance. To establish an interpretation scheme, we consider it important to take into account the uncertain nature of unstable resonance. Moreover, we would like to distinguish states having reasonably interpretable internal structure as discussed in Sec.~\ref{sec:eigenstates}. Motivated by these ideas, in this section, we propose an interpretation scheme for the complex compositeness of resonances. 

\subsection{Measurements}
\label{subsec:measurement} 

Before introducing the interpretation scheme, we first focus on the fact that the expectation values associated with unstable states generally become complex. We discuss how this complex nature is related to the instability of the states by referring to Ref.~\cite{Berggren:1970wto}. In Ref.~\cite{Berggren:1970wto}, the author discusses the extraction of the resonance contribution from a transition process in nuclear reactions represented by a complex expectation value. It is proposed that the intermediate components of the reaction are classified into the following three categories:
\begin{itemize}
\item[(i)] practically certain identification as $\ket{{\rm resonance}}$;
\item[(ii)] practically certain identification as not $\ket{{\rm resonance}}$;
\item[(iii)] uncertain whether $\ket{{\rm resonance}}$ or not.
\end{itemize}
It is unique in Ref.~\cite{Berggren:1970wto} to introduce the third category. This additional category (iii) represents the ambiguous properties of resonances; in extracting the resonance contribution from the spectrum, the eigenenergy is not unique due to the finite lifetime, and the separation of the resonance from the background is to some extent ambiguous. It is proposed in Ref.~\cite{Berggren:1970wto} to express the probabilities $a$, $b$, and $c$ for cases (i), (ii), and (iii), respectively, in terms of the complex matrix element $p$ as
\begin{align}
|p|&=a+c, \quad
|1-p|= b+c,
\label{eq:Berggren}
\end{align}

\begin{figure}[t]
\centering
\includegraphics[width=0.6\textwidth]{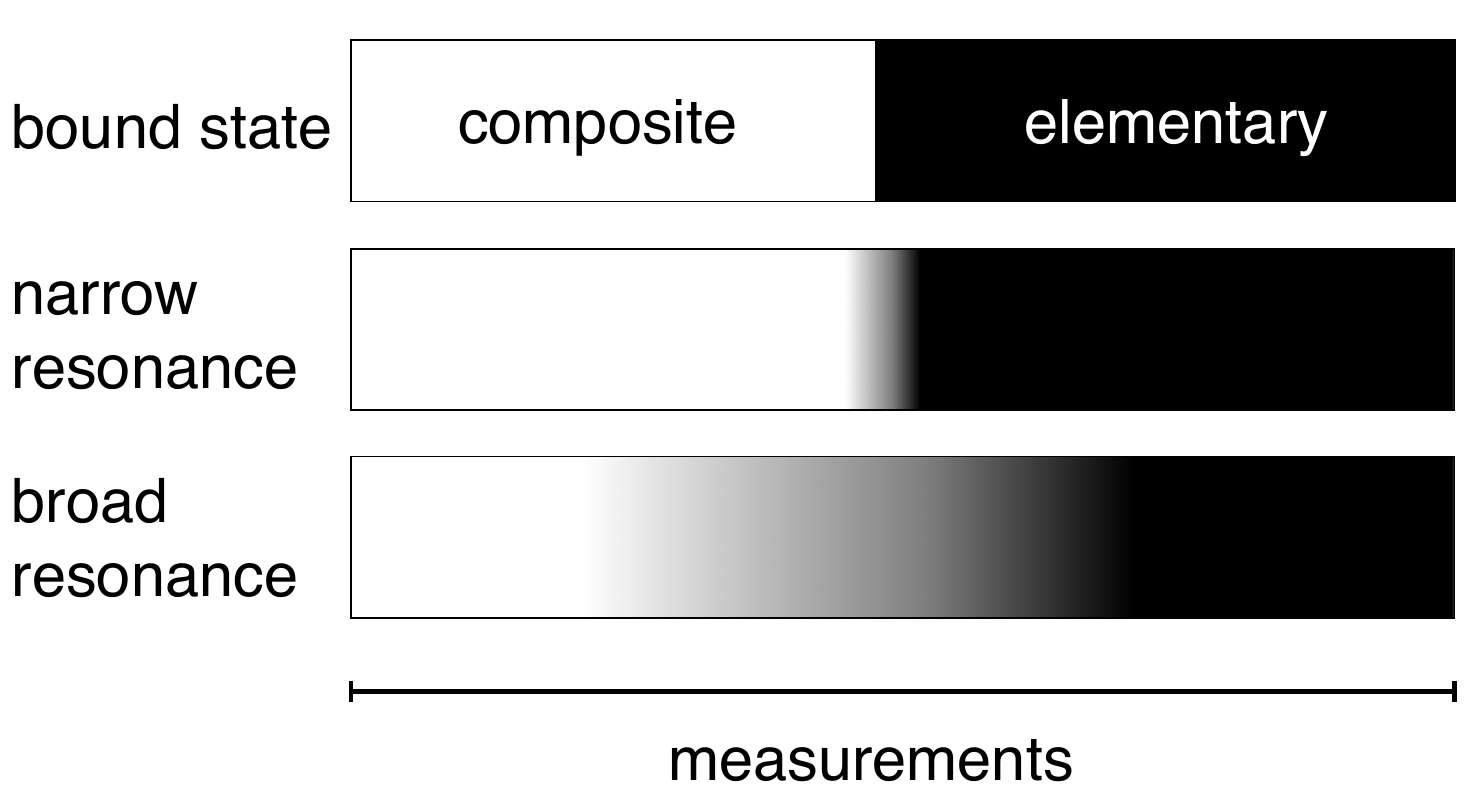}
\caption{The schematic illustration of the interpretation of compositeness $X$ and elementarity $Z$ of a stable bound state (top row), a narrow resonance (middle row), and a broad resonance (bottom row). The results of the projective measurements of the internal structure of the state are represented by white (composite), black (elementary), and gray (uncertain).}
\label{fig:bound_state}
\end{figure}

From now on, we consider the complex compositeness based on the idea in Ref.~\cite{Berggren:1970wto} that the complex expectation value reflects the uncertainty of the measurement due to the unstable nature of the resonance. To start with, we recall how the compositeness $X$ of a stable bound state is interpreted as the probability. The wavefunction of the bound state $\ket{B}$ is expressed by the linear combination $\ket{B}=\sqrt{X}\ket{\rm composite}+\sqrt{Z}\ket{\rm elementary}$. The compositeness $X=\braket{B|{\rm composite}}\braket{{\rm composite}|B}$ can then be regarded as the expectation value of the projection operator $\ket{\rm composite}\bra{\rm composite}$~\cite{Oller:2017alp,Kinugawa:2024crb}. Namely, $X$ is obtained by performing many projective ``measurements'' of the internal structure of the bound state $\ket{B}$. The result of a single measurement gives either 1 (the state is identified as $\ket{\rm composite}$) or 0 (identified as $\ket{\rm elementary}$), and the probability of finding $1$ in many measurements is the compositeness $X$, which takes the value $0\leq X \leq 1$. This is schematically illustrated in Fig.~\ref{fig:bound_state} (top row), where we sort out the results of measurements from left to right by representing the measurement outcome 1 (0) in white (black). In this case, $X$ and $Z$ correspond to the fractions of white and black regions, respectively. 

For an unstable resonance, the wavefunction can be expressed by the linear combination $\ket{R}=\sqrt{X}\ket{\rm composite}+\sqrt{Z}\ket{\rm elementary}$. Due to the unstable nature of the resonance, the expectation value should be taken with the Gamow vector $\bra{\tilde{R}}$, leading to the complex valued compositeness~\cite{Kamiya:2015aea,Kamiya:2016oao,Kukulin,Kinugawa:2024crb}
\begin{align}
X=\braket{\tilde{R}|{\rm composite}}\braket{{\rm composite}|R} \in \mathbb{C}.
\end{align}
Here, we discuss the interpretation of this complex number from the perspective of the measurements using the projection operator. We regard that the complex expectation value $X$ reflects the uncertainty of the measurement due to the unstable nature of the resonance, as suggested in Ref.~\cite{Berggren:1970wto}. This uncertainty would imply that the result of the measurement cannot be classified as unambiguously composite (white) or elementary (black), as in bound state cases. Among the results that cannot be classified unambiguously, some may exhibit a composite-like nature. In this case, we express the degree of composite nature of the resonance in a single measurement by the grayscale. We sort out the results of the measurements according to the composite nature from left to right, in Fig.~\ref{fig:bound_state} (middle and bottom rows). For a narrow resonance (middle row), the uncertainty of the measurement is expected to be small and most of the measurements can be well classified into composite or elementary. On the other hand, a broad resonance (bottom row) contains a substantial gray region, which is uncertain whether composite or elementary. 

\subsection{Interpretation scheme}
\label{subsec:XYZ}

By taking into account the relation between the uncertain nature of resonances and the measurement of the internal structure discussed in the previous section, here we consider the interpretation of the complex compositeness. In the previous proposals~\cite{Aceti:2012dd,Xiao:2012vv,Aceti:2014ala,Hyodo:2013iga,Hyodo:2013nka,Sekihara:2013sma,Kamiya:2015aea,Sekihara:2015gvw,Kamiya:2016oao,Matuschek:2020gqe}, the identification of the internal structure of resonances is classified into two categories: finding $\ket{\rm composite}$ and finding $\ket{\rm elementary}$. The probability of finding $\ket{\rm composite}$ is regarded as the compositeness of resonance. In this work, we replace these categories with ``certainly'' finding $\ket{\rm composite}$ and ``certainly'' finding $\ket{\rm elementary}$ by taking into account the uncertainty of resonances. Furthermore, we introduce an additional category ``uncertain identification'', inspired by the category (iii) mentioned above. Consequently, we propose to introduce the three probabilities $\mathcal{X}$, $\mathcal{Y}$, and $\mathcal{Z}$:
\begin{itemize}
\item[$\mathcal{X}$] : probability of certainly finding $\ket{{\rm composite}}$;
\item[$\mathcal{Y}$] : probability of uncertain identification;
\item[$\mathcal{Z}$] : probability of certainly finding $\ket{{\rm elementary}}$.
\end{itemize}

Now we relate $\mathcal{X,Y,Z}$ with the complex compositeness $X$. For a sensible probabilistic interpretation, $\mathcal{X,Y,Z}$ should be normalized to unity. Furthermore, it is natural to expect $\mathcal{Y}=0$ for bound states because $\mathcal{Y}$ originates in the uncertain nature of resonances. For $\mathcal{X,Y,Z}$ to serve as the natural extension of $X$ and $Z$ of bound states, $\mathcal{X}$ should be a probabilistic quantity for bound states. Therefore, we impose the following two conditions for $\mathcal{X,Y,Z}$;
\begin{itemize}
\item[(I)] normalization condition: $\mathcal{X}+\mathcal{Y}+\mathcal{Z}=1$,
\item[(II)] bound state condition: $\mathcal{X}\to X$, $\mathcal{Z}\to Z$, and $\mathcal{Y}\to 0$ for bound states ($0\leq X\leq 1$).
\end{itemize}

In what follows, we use the measurement picture illustrated in Fig.~\ref{fig:bound_state} to construct relations between the probabilities $\mathcal{X}$, $\mathcal{Y}$, and $\mathcal{Z}$ and the complex compositeness $X$ that satisfies the two conditions introduced above. In Fig.~\ref{fig:bound_state}, $\mathcal{X}$ and $\mathcal{Z}$ would correspond to the fractions of certainly white and certainly black regions, respectively. The fraction of the gray region in between would then represent $\mathcal{Y}$. We note that the boundary between certainly composite and uncertain identifications is not unique and should be determined by our choice. Namely, we need to determine the boundary to establish the interpretation scheme. 

We then introduce the ``possibly'' composite fraction by combining the results exhibiting a composite-like nature to some extent with those in certainly composite $\mathcal{X}$. In this case, the possibly composite (elementary) probability should contain the certainly composite (elementary) probability. We also allow that the possibly composite fraction has an overlap with the possibly elementary fraction, when some of the measurements exhibit both the composite and elementary nature. As discussed in Fig.~\ref{fig:bound_state}, a resonance with large uncertainty contains a substantial gray region, indicating the possibly composite fraction is large. This shares a common feature with the magnitudes of $|X|$ and $|Z|$, which are expected to increase for a resonance with large uncertainty. We thus identify the possibly composite and elementary fractions as $|X|$ and $|Z|$, respectively:
\begin{itemize}
\item[$|X|$] : fraction of possibly finding $\ket{{\rm composite}}$;
\item[$|Z|$] : fraction of possibly finding $\ket{{\rm elementary}}$.
\end{itemize}
Since $|X|+|Z| \geq 1$ holds from the triangle inequality, the possibly composite and possibly elementary fractions always overlap with each other, as shown in Fig.~\ref{fig:interpretation} (a). Moreover, the possibly composite (elementary) fraction should include the certainly composite (elementary) probability, and hence, we require
\begin{align}
|X|& > \mathcal{X},\quad
|Z| > \mathcal{Z}.
\end{align}
The difference between $|X|$ (possibly composite) and $\mathcal{X}$ (certainly composite) reflects the uncertain nature of the resonance. It is therefore expected that this difference increases for an uncertain resonance and vanishes in the bound-state limit. This observation suggests that the difference between $|X|$ and $\mathcal{X}$ is proportional to $\mathcal{Y}$ as
\begin{align}
|X|&=\mathcal{X}+\alpha \mathcal{Y}, \quad
|Z|= \mathcal{Z}+\alpha \mathcal{Y},
\label{eq:XYZ}
\end{align}
by introducing a real parameter $\alpha>0$, which controls the weight of $\mathcal{Y}$. For $\alpha = 1$, this assignment is similar to Eq.~\eqref{eq:Berggren} proposed in Ref.~\cite{Berggren:1970wto}. We will discuss the role of $\alpha$ in Sec.~\ref{subsec:alpha}. We note that the relation between $\mathcal{X},\mathcal{Y},\mathcal{Z}$ and $|X|,|Z|$ is not uniquely determined, and the present prescription is one reasonable choice. With the assignment~\eqref{eq:XYZ} and the normalization condition $\mathcal{X}+\mathcal{Y}+\mathcal{Z}=1$, we obtain
\begin{align}
\mathcal{X}&=\frac{(\alpha-1)|X|-\alpha|Z|+\alpha}{2\alpha-1}, \label{eq:calX} \\
\mathcal{Y}&=\frac{|X|+|Z|-1}{2\alpha-1}, \label{eq:calY} \\
\mathcal{Z}&=\frac{(\alpha-1)|Z|-\alpha|X|+\alpha}{2\alpha-1}. \label{eq:calZ}
\end{align}

To proceed, we recall that the probabilities $\mathcal{X,Y,Z}$ should be positive. In this work, we impose the condition $\alpha>1/2$ so that $\mathcal{Y}$ is always positive due to the triangle inequality. Even in this case, $\mathcal{X}$ and $\mathcal{Z}$ can be negative, for example, when $\mathcal{Y}$ is substantially large. This case should be associated with large $|{\rm Im}\ X|$, indicating that the resonance exhibits a highly uncertain nature, based on the definition of the probability $\mathcal{Y}$. This suggests that $\mathcal{X}<0$ or $\mathcal{Z}<0$ occurs for highly uncertain states with a large width for which a sensible interpretation of the internal structure is not possible, as discussed in Sec.~\ref{sec:eigenstates}. We thus adopt a criterion that the internal structure of resonances with $\mathcal{X}<0$ or $\mathcal{Z}<0$ cannot be quantified by the compositeness.

In summary, we propose the following classification of resonances (Table~\ref{tab:sum}). We first fix the value of $\alpha$, and then evaluate $\mathcal{X,Y,Z}$ from complex $X$. If either $\mathcal{X}$ or $\mathcal{Z}$ is negative, we classify the resonance as a non-interpretable state due to the uncertain nature. Else, if both $\mathcal{X}$ and $\mathcal{Z}$ are positive, the internal structure of the resonance can be interpreted in the present scheme. If $\mathcal{X}$ ($\mathcal{Z}$) is the largest, the state is dominated by the composite (elementary) component. If $\mathcal{Y}$ is the largest, the nature of the state is uncertain, whether elementary or composite. In contrast to the previous proposals, we restrict ourselves to interpreting the resonances that share a similar feature with the bound states, and classify the other states as uncertain or non-interpretable. We note that this classification is one possible prescription for interpreting the complex compositeness and depends on the parameter $\alpha$, as discussed in Sec.~\ref{subsec:choice-alpha}.

 \begin{table}
 \centering
 \caption{Classification of resonances with $\mathcal{X,Y,Z}$.\label{tab:sum}}
  \begin{tabular}{ccc} \hline \hline
    &\ $\mathcal{X}> \mathcal{Y}$ and $\mathcal{X}>\mathcal{Z}$ &\ composite dominant \\
   $\mathcal{X}\geq 0$ and $\mathcal{Z}\geq 0$  &\ $\mathcal{Z}> \mathcal{Y}$ and $\mathcal{Z}>\mathcal{X}$ &\ elementary dominant\\
    &\ $\mathcal{Y}> \mathcal{X}$ and $\mathcal{Y}>\mathcal{Z}$ &\ uncertain\\
     \hline
    $\mathcal{X}< 0$ or $\mathcal{Z}< 0$ &  & non-interpretable \\
      \hline \hline
  \end{tabular}
 \end{table} 

\subsection{Interpretation scheme with parameter $\alpha$}
\label{subsec:alpha}

\begin{figure}[t]
\centering
\includegraphics[width=0.45\textwidth]{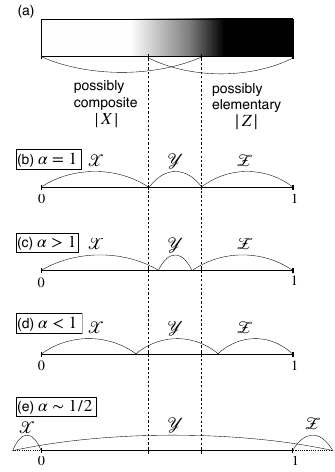}
\caption{The schematic illustration of the interpretation of complex compositeness $X$ and elementarity $Z$ of an unstable resonance.}
\label{fig:interpretation}
\end{figure}

From now on, we discuss the role of $\alpha$ by studying the $\alpha$ dependence of the interpretation scheme. From Eq.~\eqref{eq:XYZ}, $\alpha \mathcal{Y}$ determines the difference between the possibly composite probability $|X|$ and the certainly composite probability $\mathcal{X}$. Namely, $\alpha$ determines the weight of $\mathcal{Y}$ in this equation. In other words, the magnitude of the certainly composite probability $\mathcal{X}$ is controlled by $\alpha$ for a fixed $X\in\mathbb{C}$. Alternatively, we may regard that the numerator of $\mathcal{Y}$ in Eq.~\eqref{eq:calY} $(|X|+|Z|-1)$ corresponds to the overlap of possibly composite and possibly elementary components, and $\alpha$ in the denominator determines the weight of the overlap in $\mathcal{Y}$. In this way, the boundary between $\mathcal{X}$ and $\mathcal{Y}$ is determined by the parameter $\alpha$ (Fig.~\ref{fig:interpretation}).

Let us discuss the probabilities $\mathcal{X,Y,Z}$ with particular choices of $\alpha$. Panels (b), (c), and (d) in Fig.~\ref{fig:interpretation} show the correspondence of the measurements to the probabilities $\mathcal{X,Y,Z}$ with $\alpha=1$, $\alpha>1$, and $\alpha<1$ cases, respectively. In the case with $\alpha=1$ [Fig.~\ref{fig:interpretation} (b)], the uncertain probability $\mathcal{Y}$ corresponds to the overlap of the possibly composite and elementary probabilities, $|X|+|Z|-1$. In other words, the difference between the possibly composite probability ($|X|$) and the certainly composite probability $\mathcal{X}$ is $\mathcal{Y}$. Equivalently, $\mathcal{Z}$ (certainly elementary) is identified as $1-|X|$ (complement of possibly composite fraction).

For $\alpha>1$ [Fig.~\ref{fig:interpretation} (c)], Eq.~\eqref{eq:calY} shows that the probability $\mathcal{Y}$ decreases and therefore $\mathcal{X}$ and $\mathcal{Z}$ increase from those with $\alpha=1$. This means that we classify a larger fraction of the measurements into $\mathcal{X}$ and $\mathcal{Z}$ than $\alpha=1$. From Eq.~\eqref{eq:calY}, for a larger $\alpha$, the probability $\mathcal{Y}$ becomes smaller. In the $\alpha\to \infty$ limit, we obtain
\begin{align}
\mathcal{X} &= \frac{(1 - 1/\alpha)|X| - |Z| +1}{2 - 1/\alpha} \to 
 \frac{|X| - |Z| +1}{2}
=\tilde{X}_{\rm KH},\\
\mathcal{Z} &= \frac{(1 - 1/\alpha)|Z| - |X| +1}{2 - 1/\alpha} \to 
\frac{|Z| - |X| +1}{2}
=\tilde{Z}_{\rm KH},\\
\mathcal{Y} &= \frac{(|X| + |Z| - 1)/\alpha}{2 - 1/\alpha} \to 0 ,
\end{align}
where $\mathcal{Y}$ vanishes, and all the measurements are classified into either certainly composite $\mathcal{X}$ or certainly elementary $\mathcal{Z}$. In addition, $\mathcal{X}$ and $\mathcal{Z}$ in the $\alpha\to \infty$ limit reduce to $\tilde{X}_{\rm KH}$ and $\tilde{Z}_{\rm KH}$ proposed in Refs.~\cite{Kamiya:2015aea,Kamiya:2016oao}. Because $0 \leq \tilde{X}_{\rm KH} \leq 1$ and $0 \leq \tilde{Z}_{\rm KH} \leq 1$ for any $X \in \mathbb{C}$, resonances are always interpretable for $\alpha \to \infty$.  

If $\alpha<1$ [Fig.~\ref{fig:interpretation} (d)], the probability $\mathcal{Y}$ increases from that with $\alpha=1$. This corresponds to assigning uncertain identification $\mathcal{Y}$ to those classified into $\mathcal{X}$ and $\mathcal{Z}$ with $\alpha=1$. When $\alpha \sim 1/2$, $\mathcal{Y}$ can be larger than unity. In this case, $\mathcal{X}$ and $\mathcal{Z}$ become negative [Fig.~\ref{fig:interpretation} (e)] due to the normalization $\mathcal{X}+\mathcal{Y}+\mathcal{Z}=1$. As we discussed above, we regard the state with negative $\mathcal{X}$ or $\mathcal{Z}$ as non-interpretable. In the limit $\alpha \to 1/2$, the probabilities $\mathcal{X,Y,Z}$ diverge, corresponding to the non-interpretable assignment for any states except for the bound state satisfying $0\leq X\leq 1$. 

\subsection{Choice of $\alpha$}
\label{subsec:choice-alpha}

The interpretation scheme introduced in the previous sections contains a free parameter $\alpha$. In this section, we first examine how the classification of a given complex compositeness $X$ depends on $\alpha$. We then choose $\alpha$ for the following analysis of near-threshold $s$-wave resonances in the ERE.

\begin{figure}[t]
\centering
\includegraphics[width=0.3\textwidth]{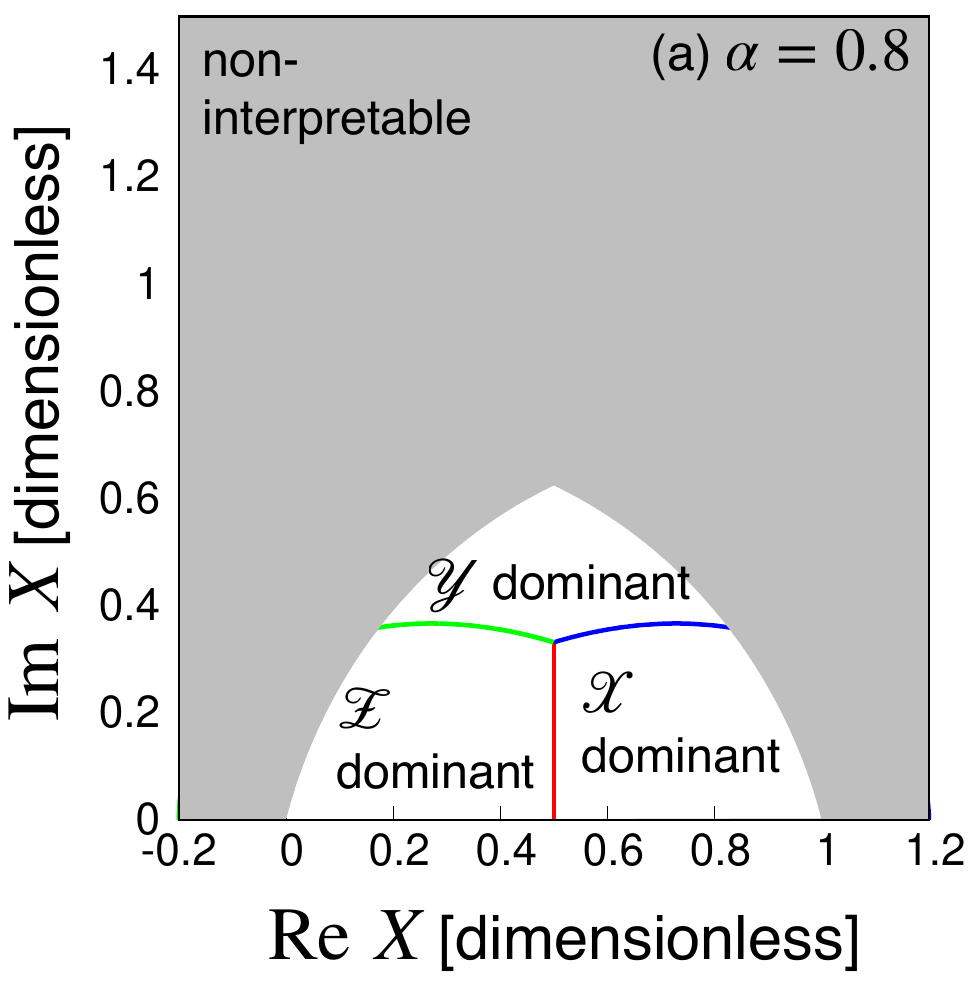}
\includegraphics[width=0.3\textwidth]{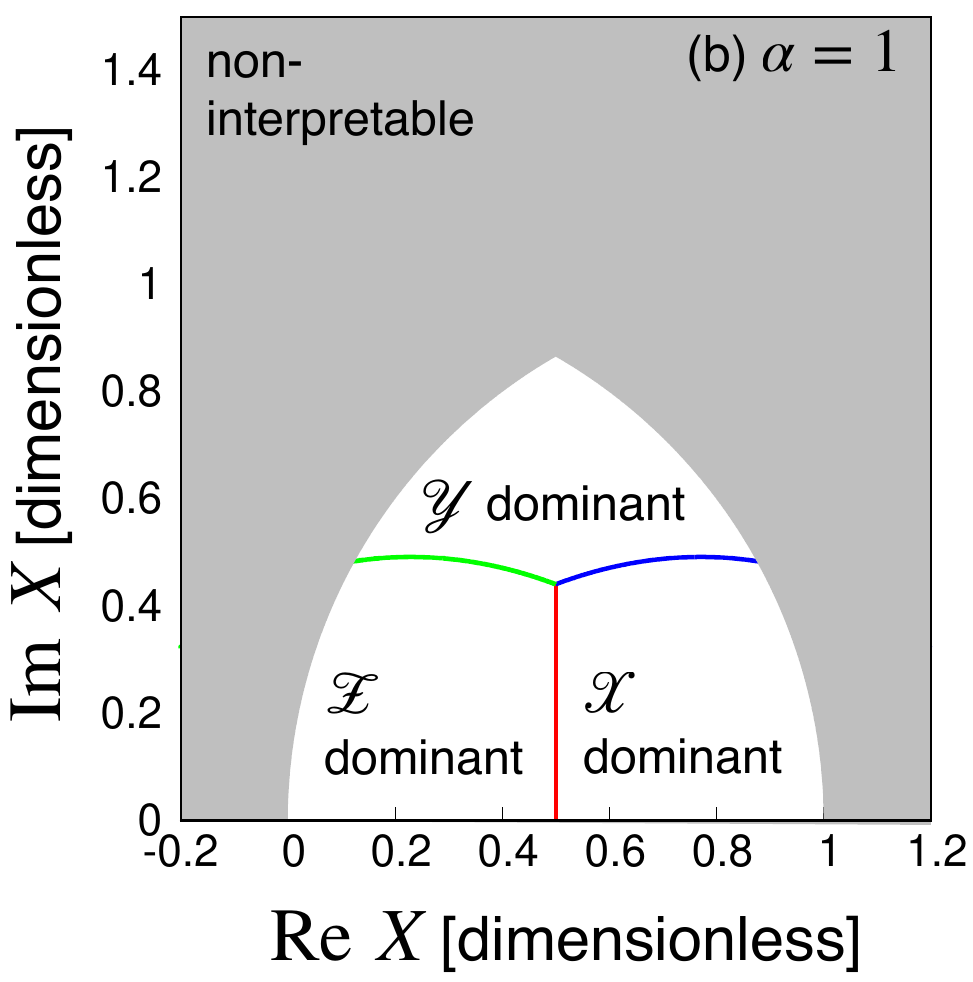}
\includegraphics[width=0.3\textwidth]{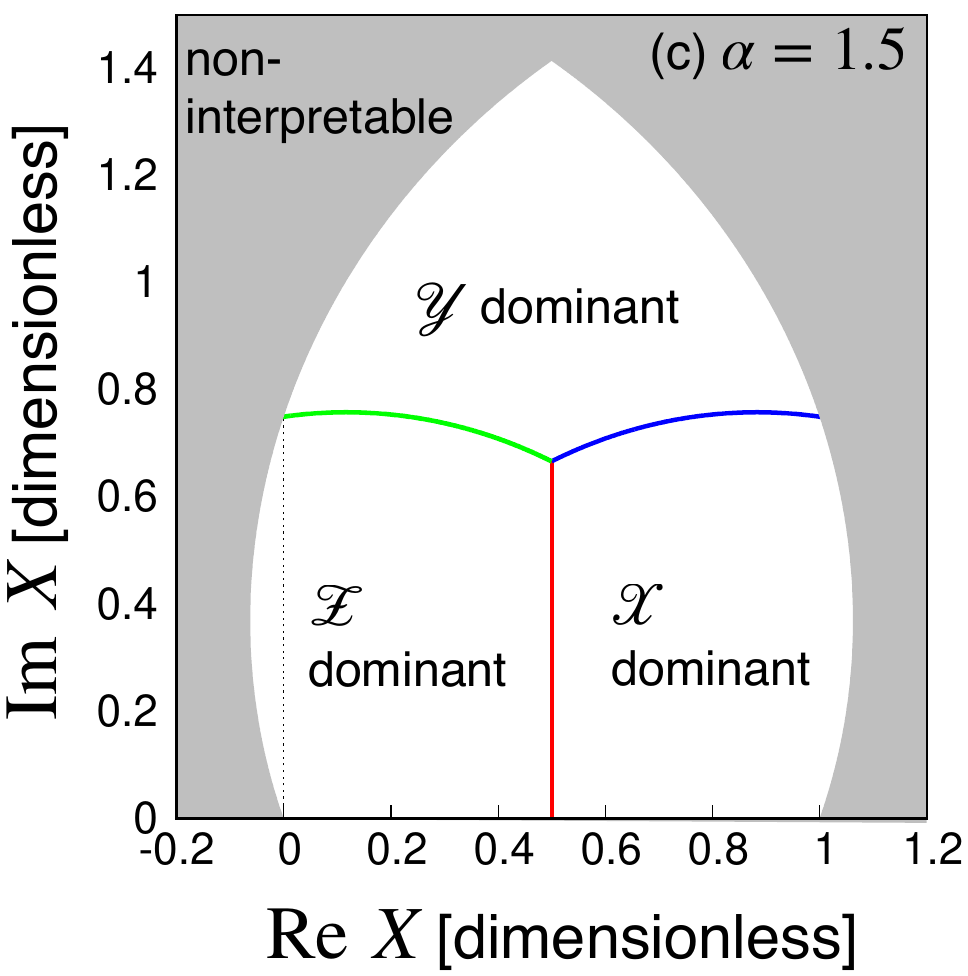}
\caption{The $\mathcal{X,Y,Z}$ dominant regions and non-interpretable region (shaded area) in the complex compositeness $X$ plane with $\alpha = 0.8$, 1, and 1.5.}
\label{fig:alphas}
\end{figure}

To examine $\alpha$ dependence, we classify the complex $X$ plane according to the interpretation scheme introduced above. At each point in the complex compositeness $X$ plane, we evaluate the corresponding quantities $\mathcal{X,Y,Z}$ and assign the point to the dominant region of the largest component. If one of the components becomes negative, the point is classified as non-interpretable. The resulting classification into the $\mathcal{X,Y,Z}$ dominant regions and the non-interpretable region is shown in Fig.~\ref{fig:alphas} for $\alpha = 0.8$, 1, and 1.5. Because $|X|$ and $|Z|=|1-X|$ remain unchanged under Im $X \to -\Im \ X$, the classification of the complex $X$ plane is symmetric with respect to the real axis, $X\to X^{*}$. In other words, it is sufficient to classify the upper half of the complex $X$ plane.

We first focus on the common feature in the three panels in Fig.~\ref{fig:alphas}. On the real axis, the bound states with $1/2<X\leq 1$ ($0\leq X<1/2$) are composite (elementary) dominant by definition. In the region near the real axis, corresponding to unstable resonances, the states with $\Re X>1/2$ ($\Re X<1/2$) are composite (elementary) dominant. This feature ensures that our proposal serves as the natural extension of the compositeness of the bound state. Above the composite-dominant and elementary-dominant regions, there is a region where the uncertain probability $\mathcal{Y}$ becomes the dominant one. This reflects the idea that $\mathcal{Y}$ characterizes the uncertain nature of the resonance, which is expected to increase for the states with large Im~$X$. If the complex $X$ exists sufficiently far from the $0 \leq X \leq 1$ region, the state is classified as non-interpretable. This is consistent with our expectation that such states are uncertain. 

We then discuss the $\alpha$ dependence in Fig.~\ref{fig:alphas}. The interpretable region with $\alpha = 1.5$ ($\alpha = 0.8$) is larger (smaller) than that of $\alpha = 1$. This tendency is consistent with the fact that the whole region becomes interpretable in the $\alpha\to \infty$ limit, while only the $0\leq X\leq 1$ region is interpretable in the $\alpha\to 1/2$ limit. In this way, the parameter $\alpha$ determines the boundary of the interpretable region.

After examining the $\alpha$ dependence of the classification, we now discuss the choice of the value of $\alpha$ by focusing on the eigenenergy of resonances $E=E_{R}-i\Gamma/2$. We regard resonances with a large $\Gamma/\Re\, E$ as states whose internal structure is non-interpretable. This is because broad resonances do not exhibit localization of the wavefunction near the origin, and hence lack the spatially localized object required for the compositeness to quantify their internal structure, as discussed in Sec.~\ref{sec:eigenstates}.
Based on this idea, we propose to classify resonances with $\Gamma> E_{R} $ as non-interpretable, by using the excitation energy $E_{R}$ as the characteristic scale. Namely, we define the boundary between interpretable and non-interpretable states by
\begin{align}
\Re E = -2\Im E,\quad \theta_{E}=\arctan\left(-\frac{1}{2}\right) ,
\label{eq:interpretable}
\end{align}
in the complex $E$ plane. Using the relation in Eq.~\eqref{eq:X-ERE-theta} between the argument of the eigenenergy $E$ and the compositeness $X$ in the ERE, the boundary~\eqref{eq:interpretable} is straightforwardly implemented as a condition for $\alpha$. Because $\mathcal{X}$ and $\mathcal{Z}$ depend on $\alpha$ as in Eqs.~\eqref{eq:calX} and \eqref{eq:calZ}, the boundary between the non-interpretable and interpretable regions, determined by the values of $\mathcal{X}$ and $\mathcal{Z}$, also depends on $\alpha$. From these observations, we define $\alpha_{0}$ as the value of $\alpha$ with which Eq.~\eqref{eq:interpretable} gives the boundary. $\alpha_{0}$ is then determined by the condition $\mathcal{X}=0$ at $\theta_{E}$ as (see Appendix~\ref{sec:alpha-calc})
\begin{align}
\alpha_{0}&=\frac{\sqrt{5}-1+\sqrt{10-4\sqrt{5}}}{2}\approx 1.1318.
\label{eq:alpha-0}
\end{align}
In the following, we set $\alpha=\alpha_{0}$.

\section{Internal structure of near-threshold resonances}
\label{sec:structure}

\begin{figure}[t]
\centering
\includegraphics[width=0.45\textwidth]{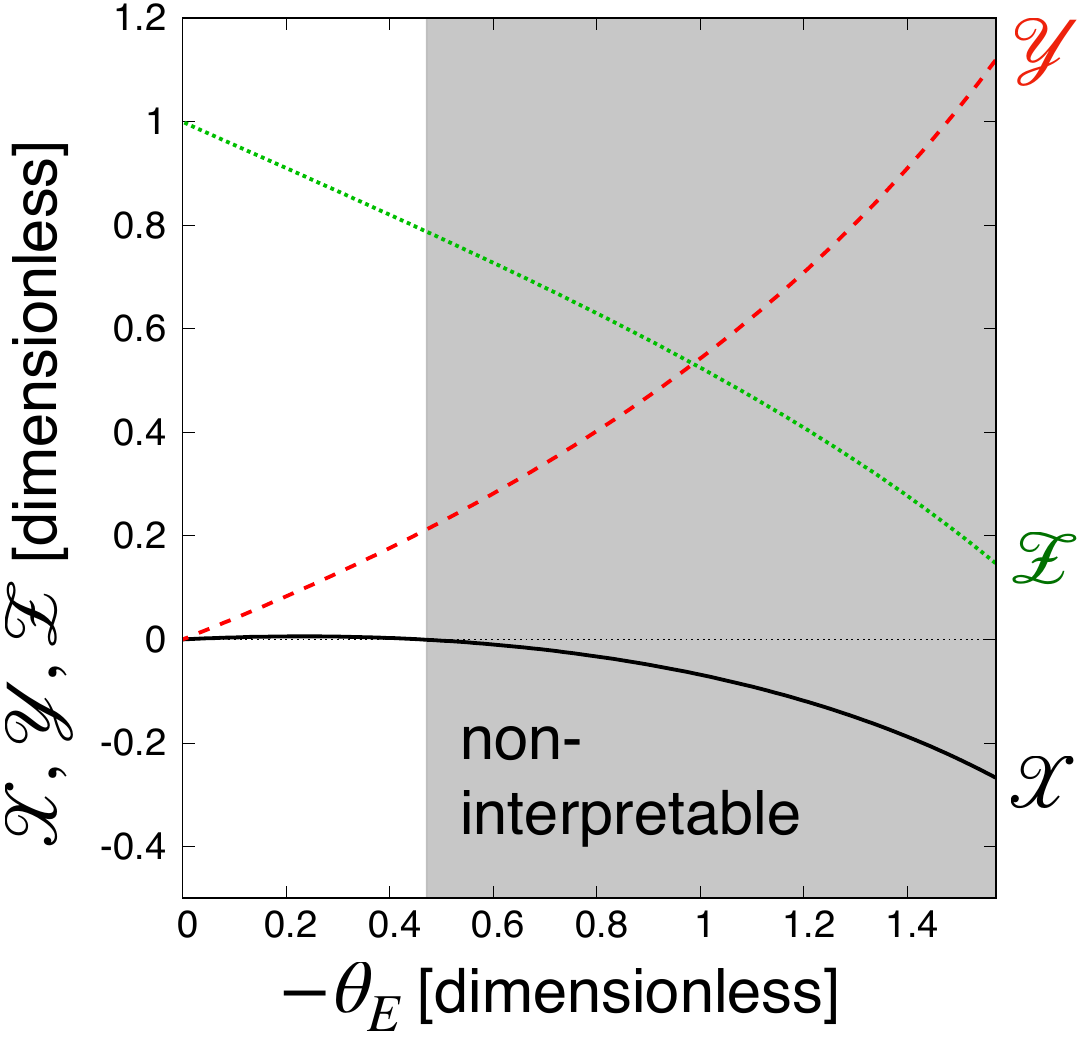}
\caption{The probabilities $\mathcal{X}$, $\mathcal{Y}$, and $\mathcal{Z}$ as functions of the argument of the eigenenergy $-\theta_{E}$. The solid line stands for $\mathcal{X}$, dashed for $\mathcal{Y}$, and dotted for $\mathcal{Z}$. The shaded area is the non-interpretable region determined by the condition~\eqref{eq:interpretable}.}
\label{fig:XYZ-theta-E}
\end{figure}

Having established the interpretation scheme, we numerically evaluate the probabilities $\mathcal{X,Y,Z}$ of near-threshold $s$-wave resonances where $|a_{0}|$ and $|r_{e}|$ are much larger than other length scales in the system. Because the compositeness of resonances in the ERE in Eq.~\eqref{eq:X-ERE-theta} depends only on the argument $\theta_{E}$, we plot $\mathcal{X,Y,Z}$ as functions of $\theta_{E}$ in the $0< -\theta_{E}\leq \pi/2$ region in Fig.~\ref{fig:XYZ-theta-E}. Note that $\mathcal{X}$ is positive for $-\theta_{E} < |\arctan(-1/2)|\approx 0.464$, although the magnitude is very small.

Let us focus on the interpretable region $0<-\theta_{E}<|\arctan(-1/2)|$. In the $\theta_{E}\to 0$ limit, $\mathcal{X}=\mathcal{Y}= 0$ and $\mathcal{Z}= 1$ because the complex $X$ goes to zero as seen in Eq.~\eqref{eq:X-ERE-theta}. The pole on the real axis with $\theta_{E}\to 0$ represents the state which does not couple to the scattering states with $\Gamma/\Re\, E = 0$. Because the coupling to the scattering states induces the compositeness of the state, it is natural to obtain the completely elementary state with $\mathcal{Z}=1$ (e.g., the bare state). The value of $\mathcal{Z}$ decreases from $\mathcal{Z}=1$ as $-\theta_{E}$ increases, but the resonance in the interpretable region remains elementary dominant with $\mathcal{Z}\gtrsim 0.8$. We therefore conclude that near-threshold narrow interpretable resonances with small values of $-\theta_{E}$ are elementary dominant in accordance with the model analysis in Ref.~\cite{Hyodo:2013iga}. This result shows that the property of interpretable near-threshold resonances is completely opposite to near-threshold bound states, which are shown to be composite dominant by the low-energy universality~\cite{Hyodo:2013iga,Hanhart:2014ssa,Kinugawa:2023fbf}. 

\begin{figure}[t]
\centering
\includegraphics[width=0.45\textwidth]{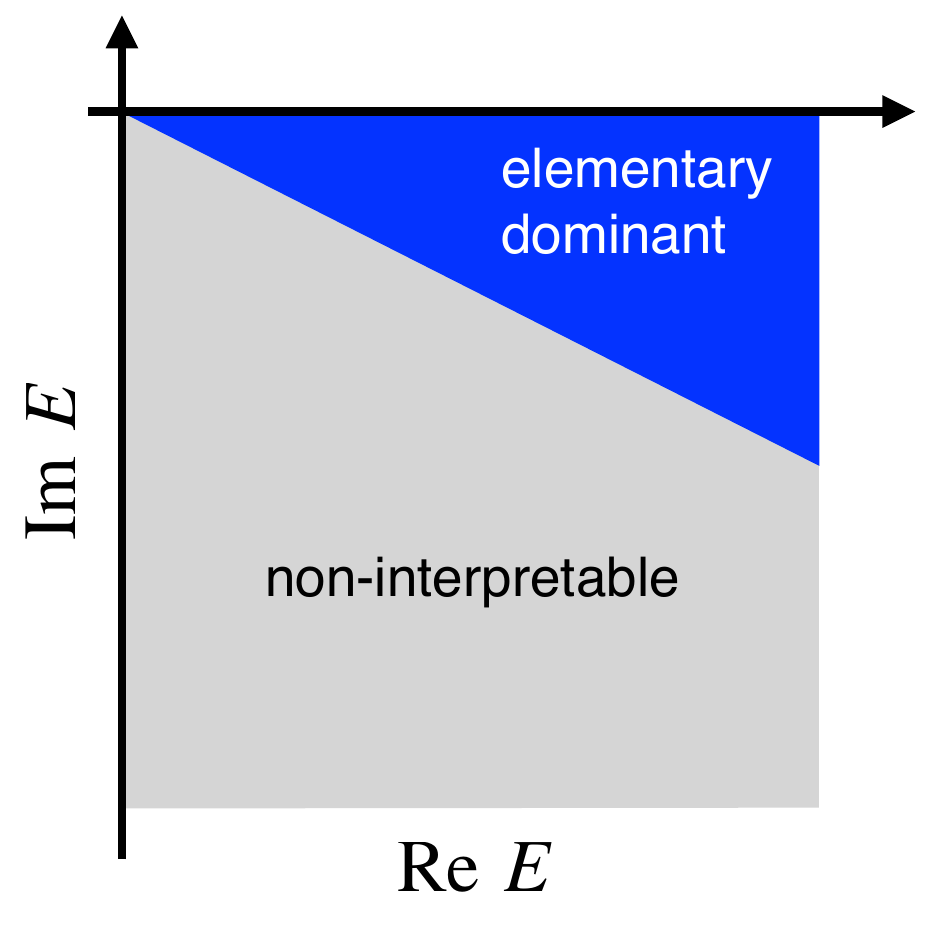}
\caption{Schematic classification of the eigenenergy of near-threshold resonances in the complex energy plane.}
\label{fig:Z-non}
\end{figure}

For the visual illustration, we show the classification of near-threshold resonances in the complex energy plane in Fig.~\ref{fig:Z-non}. We see that narrow resonances [$-\theta_{E}<|\arctan(-1/2)|$] are elementary dominant, and broad resonances are non-interpretable. Namely, in the present classification, near-threshold resonances are not classified as composite dominant. Because the ERE does not assume the origin of resonances, this conclusion holds universally for near-threshold narrow interpretable resonances. Thus, the property of near-threshold states being universally determined is common to both bound states and resonances, although the dominant component is opposite from each other.

\section{Comparison with previous works}
\label{sec:comparison}

\begin{figure}[tb]
\centering
\includegraphics[width=0.9\textwidth]{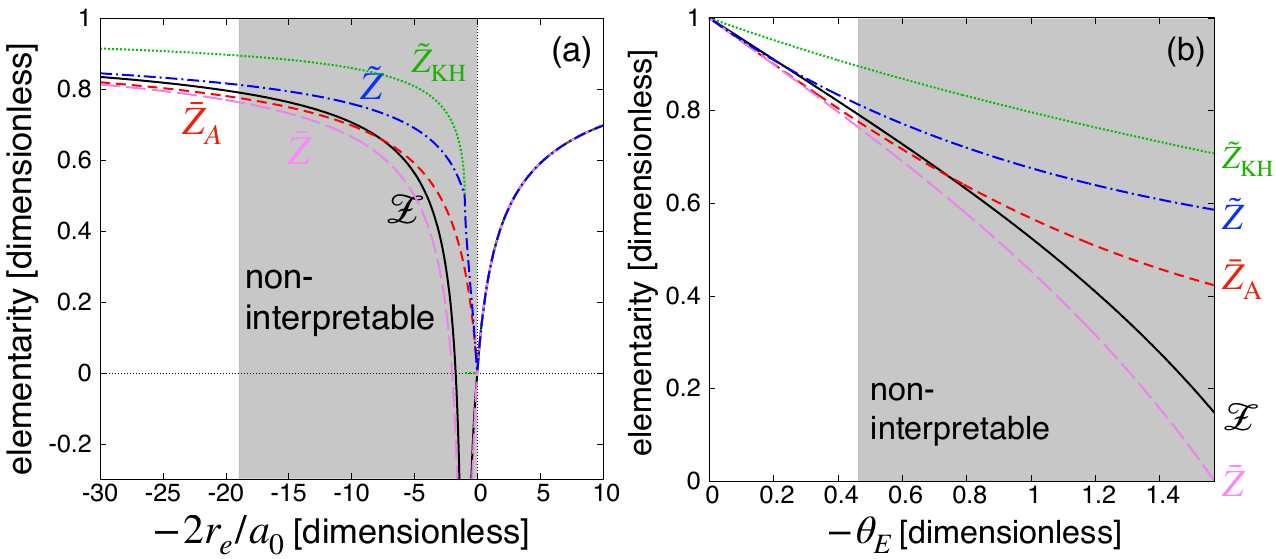}
\caption{Fractions of the elementary component as functions of the ratio $-2r_{e}/a_{0}$ [panel (a)], and those as functions of the argument of complex eigenenergy $-\theta_{E}$ [panel (b)]. The solid line stands for $\mathcal{Z}$ in this work, long dashed for $\bar{Z}$ in Ref.~\cite{Hyodo:2013iga}, dotted for $\tilde{Z}_{\rm KH}$ in Refs.~\cite{Kamiya:2015aea,Kamiya:2016oao}, dash-dotted for $\tilde{Z}$ in Ref.~\cite{Sekihara:2015gvw}, and short dashed for $\bar{Z}_{A}$ in Ref.~\cite{Matuschek:2020gqe}. The shaded areas are non-interpretable regions in this study.}
\label{fig:Z-ratio}
\end{figure}

We now compare the present results with the previous works by focusing on the elementarity of resonances. In Refs.~\cite{Hyodo:2013iga,Kamiya:2015aea,Kamiya:2016oao,Sekihara:2015gvw,Matuschek:2020gqe}, the fraction of the elementary component is defined using the scattering length $a_{0}$ and effective range $r_{e}$, or complex compositeness $X$ as follows:
\begin{align}
\bar{Z}&=1-\sqrt{\left|\dfrac{1}{1-2r_{e}/a_{0}}\right|}\quad (\text{Ref.~\cite{Hyodo:2013iga}}),\label{eq:bar-Z}\\
\tilde{Z}_{\rm KH}&=\dfrac{1-|X|+|1-X|}{2}\quad (\text{Refs.~\cite{Kamiya:2015aea,Kamiya:2016oao}}),\label{eq:tilde-Z-KH}\\
\tilde{Z}&=\dfrac{|1-X|}{|X|+|1-X|}\quad (\text{Ref.~\cite{Sekihara:2015gvw}}),\label{eq:tilde-Z}\\
\bar{Z}_{A}&=1-\sqrt{\dfrac{1}{1+|2r_{e}/a_{0}|}}\quad (\text{Ref.~\cite{Matuschek:2020gqe}}).\label{eq:Z-A}
\end{align}
These quantities reduce to the original $Z$ for bound states. Let us compare these quantities with $\mathcal{Z}$ in the present work for a given set of $a_{0}$ and $r_{e}$, where we express $X$ by $a_{0}$ and $r_{e}$ using the relation in Eq.~\eqref{eq:wbr}. We vary $1/a_{0}$ from a positive value to a sufficiently negative one for a fixed negative $r_{e}$ so that a bound state represented by the pole $k^{-}$ turns into a virtual state and then into a resonance~\cite{Hyodo:2013iga}. Because the fractions of the elementary component in Eqs.~\eqref{eq:bar-Z}, \eqref{eq:tilde-Z-KH}, \eqref{eq:tilde-Z}, \eqref{eq:Z-A} and $\mathcal{Z}$ in Eq.~\eqref{eq:calZ} depend only on the ratio of $r_{e}$ to $a_{0}$, we plot these quantities as functions of $-2r_{e}/a_{0}$ in Fig.~\ref{fig:Z-ratio}~(a). 

Before considering resonances, we discuss bound and virtual states. In the $-2r_{e}/a_{0}>0$ region where a bound state appears, all the results become identical by definition. In contrast, for a virtual state in the $-1<-2r_{e}/a_{0}<0$ region, the fractions are quite different from each other. The virtual state is interpreted as composite dominant with $\tilde{Z}_{\rm KH}$, $\tilde{Z}$, and $\bar{Z}_{A}$. In particular, with the prescription in Refs.~\cite{Kamiya:2015aea,Kamiya:2016oao}, $\tilde{Z}_{\rm KH}=0$ always holds for the virtual state, i.e., completely composite $\tilde{X}_{\rm KH}=1$. However, $\bar{Z}$ and $\mathcal{Z}$ are always negative for the virtual state. In our study, the virtual state is classified as non-interpretable, in agreement with the negative norm of the virtual state, which indicates its uncertain nature~\cite{Braaten:2007nq}. 

A resonance is represented by the pole in the $-2r_{e}/a_{0}<-2$ region in Fig.~\ref{fig:Z-ratio}~(a). The non-shaded area ($-2r_{e}/a_{0}\lesssim-18$) corresponds to the interpretable region where the resonance is narrow. To focus on the resonance, in Fig.~\ref{fig:Z-ratio}~(b), we plot the fractions as functions of the argument of the eigenenergy $-\theta_{E}$ in the whole fourth quadrant ($0< -\theta_{E}\leq \pi/2$). We see that all results converge to unity in the $\theta_{E}\to 0$ ($\Gamma/\Re\, E \to 0$) limit. Furthermore, in the interpretable region, all elementarities are larger than 0.7. In the region regarded as non-interpretable in the present study ($|\arctan{-1/2}|<-\theta_{E}$), the difference between the elementarities becomes large, and the resulting interpretations of the internal structure are not necessarily consistent. This indicates that the compositeness of broad resonances cannot be interpreted in a scheme-independent manner. Hence, it is concluded that near-threshold resonances with a narrow width are elementary dominant with any prescriptions considered here.

\section{Application to hadron resonances}
\label{sec:apply}

So far, we have discussed near-threshold resonances using the compositeness $X$ obtained from the ERE. The present interpretation scheme, however, can also be applied to hadronic states for which complex $X$ have already been evaluated by other formulations. In this section, we apply our interpretation scheme based on $\mathcal{X,Y,Z}$ to hadron resonances whose complex $X$ has been evaluated in previous studies. 

In Ref.~\cite{Kamiya:2016oao}, the complex compositeness of $\Lambda(1405)$, $f_{0}(980)$, and $a_{0}(980)$ are estimated by the weak-binding relation based on the eigenenergy and the scattering length obtained in various theoretical analyses~\cite{Ikeda:2011pi,Ikeda:2012au,Mai:2012dt,Guo:2012vv,Mai:2014xna,CDF:2011kjt,KLOE:2005jxf,Belle:2005rpz,BES:2004twe,FOCUS:2004prc,Achasov:2000ym,KLOE:2009ehb,Bugg:2008ig,Achasov:2000ku,E852:1996san}. $\Lambda(1405)$, $f_{0}(980)$, and $a_{0}(980)$ are known to be located near the threshold of $\bar{K}N$, $\bar{K}K$, and $\bar{K}K$, respectively and to decay into the $\pi\Sigma$, $\pi\pi$, and $\pi\eta$ channels. The exact pole position of each resonance depends on the analysis, and the difference among the analyses can be regarded as a theoretical uncertainty. We classify the nature of the eigenstate as a quasibound state (QB) if the real part of the eigenenergy is below the threshold, and as a resonance (R) if it is above. Because these states correspond to a below-threshold bound state and an above-threshold resonance in the limit of vanishing decay widths, we expect these states to exhibit the internal structures inferred in the present study.

We evaluate $\mathcal{X,Y,Z}$ from complex $X$ given in Ref.~\cite{Kamiya:2016oao}, and summarize the results in  Table~\ref{tab:X-Kamiya2016oao}, together with the dominant component of each estimation. Most of the results in Table~\ref{tab:X-Kamiya2016oao} indicate $\mathcal{X}$ (composite) dominant or $\mathcal{Z}$ (elementary) dominant, and there is no $\mathcal{Y}$ (uncertain) dominant classification in the cases examined. Two non-interpretable cases of $\Lambda(1405)$ show negative $\mathcal{Z}$, but its magnitude is as small as $|\mathcal{Z}|\lesssim 0.2$. These results indicate that the compositeness of $\Lambda(1405)$, $f_{0}(980)$, and $a_{0}(980)$ is interpretable within the present scheme in nearly all cases, except for the small negative value of $\mathcal{Z}$ found in the two analyses of $\Lambda(1405)$.

In Refs.~\cite{Hyodo:2014bda,Hanhart:2014ssa,Matuschek:2020gqe,Sazdjian:2022kaf,Lebed:2022vks,Hanhart:2022qxq,Kinugawa:2023fbf} as well as in the present work, it is shown that the bound state below the threshold is in general composite dominant, while the resonance above the threshold is elementary dominant. This tendency is reflected in the results of $\Lambda(1405)$ and $a_{0}(980)$ in Table~\ref{tab:X-Kamiya2016oao}; $\Lambda(1405)$ is mostly QB and composite dominant, while $a_{0}(980)$ is R and elementary dominant. The results of $f_{0}(980)$ are somehow scattered, presumably because the pole position is not yet precisely settled. In addition, we note that these states are coupled to decay channels, as mentioned above. In fact, the decay channel contribution is known to modify the value of the compositeness~\cite{Kinugawa:2023fbf}. For a more detailed discussion, therefore, we need to estimate the coupled-channel effects on the compositeness in a quantitative manner.

 \begin{table}
 \caption{Interpretation of the complex compositeness of $\Lambda(1405)$, $f_{0}(980)$, and  $a_{0}(980)$ in Ref.~\cite{Kamiya:2016oao} by $\mathcal{X,Y,Z}$. For each state, the complex compositeness $X$ of the near-threshold channel is evaluated in different analyses. In the ``Nature'' column, QB and R denote quasi-bound and resonance states, corresponding to poles located below and above the threshold, respectively, and the classification based on Table~\ref{tab:sum} is shown.
\label{tab:X-Kamiya2016oao}}
  \centering
  \begin{tabular}{cccccccc} \hline \hline
  Hadron (threshold) & Analysis & $X$ & $\mathcal{X}$ & $\mathcal{Y}$ & $\mathcal{Z}$ & Nature 
     \\ \hline 
    $\Lambda(1405)$  ($\bar{K}N$) & \cite{Ikeda:2011pi,Ikeda:2012au} & $1.2+0.1i$ & 0.82 & 0.34 & $-0.16$ & QB (non)  \\ 
    			   				  & \cite{Mai:2012dt} & $0.6+0.1i$ & 0.60 & 0.02 & 0.39 & QB ($\mathcal{X}$)  \\ 
    						      & \cite{Guo:2012vv} & $0.9-0.2i$ & 0.79 & 0.12 & 0.09 & QB ($\mathcal{X}$) \\ 
    						      & \cite{Mai:2014xna} & $0.6+0.0i$ & 0.60 & 0.00 & 0.40 & R ($\mathcal{X}$) \\ 
    						      & \cite{Mai:2014xna} & $1.0+0.5i$ & 0.56 & 0.49 & $-0.05$ & QB (non) \\ \\
    $f_{0}(980)$  ($\bar{K}K$) & \cite{CDF:2011kjt} & $0.3-0.3i$ & 0.26 & 0.15 & 0.60 & R ($\mathcal{Z}$)  \\ 
    						   & \cite{KLOE:2005jxf} & $0.3-0.1i$ & 0.30 & 0.02 & 0.69 & QB ($\mathcal{Z}$)  \\ 
    						   & \cite{Belle:2005rpz} & $0.4-0.2i$ & 0.38 & 0.06 & 0.56 & QB ($\mathcal{Z}$)  \\ 
    						   & \cite{BES:2004twe} & $0.7-0.3i$ & 0.60 & 0.15 & 0.26 & R ($\mathcal{X}$)  \\ 
    						   & \cite{FOCUS:2004prc} & $0.3-0.1i$ & 0.30 & 0.02 & 0.69 & QB ($\mathcal{Z}$) \\ 
    						   & \cite{Achasov:2000ym} & $0.9-0.2i$ & 0.79 & 0.12 & 0.09 & R ($\mathcal{X}$)  \\ \\
    $a_{0}(980)$  ($\bar{K}K$) & \cite{KLOE:2009ehb} & $0.2-0.2i$ & 0.19 & 0.09 & 0.73 & R ($\mathcal{Z}$)  \\ 
    						   & \cite{Bugg:2008ig} & $0.2-0.2i$ & 0.19 & 0.09 & 0.73 & R ($\mathcal{Z}$)  \\ 
    						   & \cite{Achasov:2000ku} & $0.8-0.4i$ & 0.59 & 0.27 & 0.14 & R ($\mathcal{X}$)  \\ 
    						   & \cite{E852:1996san} & $0.1-0.2i$ & 0.09 & 0.12 & 0.79 & R ($\mathcal{Z}$) \\ \hline \hline
  \end{tabular}
 \end{table} 

\section{Summary}
\label{sec:sum}

In this work, we examine the internal structure of near-threshold $s$-wave resonances through quantitative analysis. Initially, we demonstrate that the nature of near-threshold $s$-wave resonances is not determined solely by the scattering length, in contrast to the expectation from the low-energy universality. This is attributed to the large negative effective range characteristic of near-threshold resonances.

To analyze the internal structure of resonances, we then introduce a new probabilistic interpretation scheme for their complex compositeness. The key idea is to incorporate the probability of the uncertain identification in addition to the compositeness and elementarity. Additionally, in this framework, a discernible criterion to exclude uncertain states is inherently embedded. This is our solution related to the first issue raised in the introduction.

To address the second issue, using our interpretation scheme, we present quantitative evidence demonstrating that near-threshold interpretable resonances with a narrow decay width exhibit elementary dominance. The elementary dominance of narrow resonances holds regardless of the origin, because the compositeness and elementarity are determined within the ERE. This observation can be regarded as a kind of universality inherent to interpretable near-threshold resonances with a narrow width. Our analysis underscores the qualitative disparity between near-threshold resonances and shallow bound states, the latter being composite dominant. This indicates that care must be taken in discussing near-threshold states, as their nature crucially depends on the position relative to the threshold. 

This work holds significant implications for the analysis of exotic hadrons. While many near-threshold exotic hadrons are observed~\cite{Hosaka:2016pey,Guo:2017jvc,Brambilla:2019esw}, being near the threshold does not immediately lead to their molecular nature. In fact, in the application to hadron resonances in Sec.~\ref{sec:apply}, $\Lambda(1405)$, which is mostly classified as a quasibound state, tends to be composite dominant, whereas $a_{0}(980)$, classified as a resonance, tends to be elementary dominant. This viewpoint may also be useful for other near-threshold exotic candidates. For example, our findings indicate small $K^{-}\Lambda$ compositeness of $\Xi(1620)$ above the $K^{-}\Lambda$ threshold, aligning with the small compositeness $|g_{K^{-}\Lambda}^{2}dG/dE|=0.162$ found in Ref.~\cite{Sarti:2023wlg}.\footnote{We note that the extension of the present scheme to the coupled-channel case should be considered for more quantitative assessments.} More generally, as demonstrated in Sec.~\ref{sec:apply}, the present interpretation scheme can be applied to complex compositeness evaluated in other formulations. In this way, our interpretation scheme for the complex compositeness will be a powerful tool for elucidating the internal structure of exotic hadrons. 


\section*{Acknowledgments}
%
This work has been supported in part by the Grants-in-Aid for Scientific Research from JSPS (Grants
No.~JP26K07088, 
No.~JP26H01426, 
No.~JP25K23387, 
No.~JP23KJ1796, 
No.~JP23H05439, 
No.~JP22K03637, and 
No.~JP18H05402), 
and by JST, the establishment of university fellowships towards the creation of science technology innovation, Grant No.~JPMJFS2139. 
T.K. is supported by the RIKEN special postdoctoral researcher
program. 


\appendix
\section{$\mathcal{X,Y,Z}$ dominant regions}

\begin{figure}[t]
\centering
\includegraphics[width=0.45\textwidth]{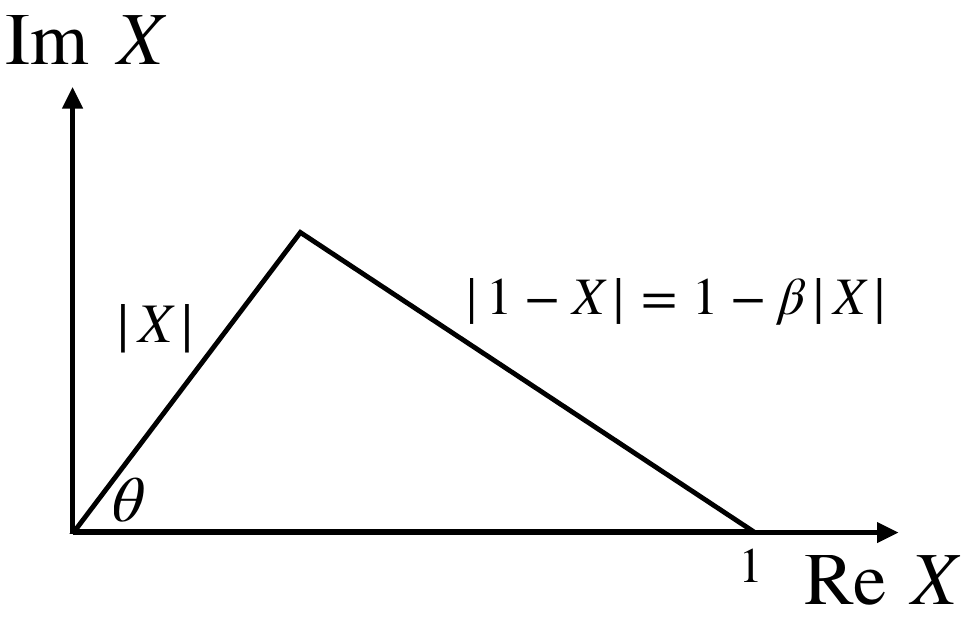}
\caption{The geometrical expression of Eq.~\eqref{eq:X=Y-triangle}.}
\label{fig:triangle}
\end{figure} 

Here, we discuss the classification of the complex compositeness $X$ into $\mathcal{X}$, $\mathcal{Y}$, $\mathcal{Z}$ dominant, or non-interpretable regions in the complex $X$ plane with fixed $\alpha$. When the complex compositeness $X$ is given, the values of $\mathcal{X,Y,Z}$ are calculated from Eqs.~\eqref{eq:calX}, \eqref{eq:calY}, and \eqref{eq:calZ}. As in Table~\ref{tab:sum}, we classify the nature of the resonance according to the values of $\mathcal{X,Y,Z}$. 

To associate each region in the complex $X$ plane with the corresponding interpretation, we consider the conditions that define the boundaries between the regions. From Eqs.~\eqref{eq:calX}, \eqref{eq:calY}, and \eqref{eq:calZ}, the boundary of the $\mathcal{X}$ dominant region and the $\mathcal{Z}$ dominant region in the $X$ plane is determined by 
\begin{align}
\mathcal{X}&=\mathcal{Z}\quad \Leftrightarrow \quad (\alpha-1)|X|-\alpha|1-X|+\alpha=(\alpha-1)|1-X|-\alpha|X|+\alpha, 
\label{eq:X=Z}
\end{align}
the boundaries of the $\mathcal{Y}$ dominant region are
\begin{align}
\mathcal{X}&=\mathcal{Y}\quad \Leftrightarrow \quad (\alpha-1)|X|-\alpha|1-X|+\alpha=|X|+|1-X|-1, 
\label{eq:X=Y}\\
\mathcal{Z}&=\mathcal{Y}\quad \Leftrightarrow \quad (\alpha-1)|1-X|-\alpha|X|+\alpha=|X|+|1-X|-1,
\label{eq:Z=Y}
\end{align}
and the boundaries of the non-interpretable region are
\begin{align}
\mathcal{X}&=0\quad \Leftrightarrow \quad 
(\alpha - 1)|X| - \alpha|1-X| + \alpha 
= 0 \label{eq:X=0} , \\
\mathcal{Z}&=0\quad \Leftrightarrow \quad 
(\alpha - 1)|1-X| - \alpha|X| + \alpha 
= 0 \label{eq:Z=0} .
\end{align}

Because Eq.~\eqref{eq:X=Z} gives $|X|=|1-X|$ for $\alpha\neq 1/2$, the boundary between $\mathcal{X}$ dominant and $\mathcal{Z}$ dominant regions is obtained as
\begin{align}
\Re\, X &= \frac{1}{2} , 
\label{eq:reX1/2}
\end{align}
which is independent of $\alpha$. This reduces to $X=1/2$ for real $X$, indicating that $\mathcal{X}$ and $\mathcal{Z}$ are natural extensions to the real-valued $X$ and $Z$. To determine the boundary between $\mathcal{X}$ dominant and $\mathcal{Y}$ dominant regions, we rewrite Eq.~\eqref{eq:X=Y} as
\begin{align}
|1-X|&=1-\beta|X|, 
\label{eq:X=Y-triangle}
\end{align}
with
\begin{align}
\beta=\frac{2-\alpha}{\alpha+1}.
\end{align}
Because $\alpha >1/2$, $\beta$ is always smaller than unity. A complex $X$ can be expressed by $X = |X|e^{i\theta}$ with the argument $\theta$. From the geometric relation shown in Fig.~\ref{fig:triangle} with Eq.~\eqref{eq:X=Y-triangle}, the cosine formula gives
\begin{align}
\cos\theta =\frac{|X|^{2}+1-(1-\beta|X|)^{2}}{2|X|},
\end{align}
which gives the boundary of $\mathcal{X}=\mathcal{Y}$ as
\begin{align}
|X|&=\frac{2\cos\theta-2\beta}{1-\beta^{2}},
\label{eq:X=Y-|X|} 
\end{align}
with $\cos\theta>\beta$. This boundary can also be expressed by the real and imaginary parts of $X$:
\begin{align}
\Re\ X&=\frac{2\cos\theta-2\beta}{1-\beta^{2}}\cos\theta, \label{eq:re-X=Y}\\
\Im\ X&=\frac{2\cos\theta-2\beta}{1-\beta^{2}}\sin \theta. \label{eq:im-X=Y}
\end{align}
These functions are known as ``lima\c{c}on of Pascal''. From the expressions of $\mathcal{X,Y,Z}$ with $X,Z$ in Eqs.~\eqref{eq:calX}, \eqref{eq:calY}, and \eqref{eq:calZ}, we find $\mathcal{X}\to \mathcal{Z}$, $\mathcal{Z}\to \mathcal{X}$, and $\mathcal{Y}\to \mathcal{Y}$ under the interchange $X\leftrightarrow Z$. This is equivalent to $X\to 1-X$ in the complex $X$ plane, i.e., the real and imaginary parts of $X$ change as $\Re\ X \to 1 - \Re\ X$, and $\Im\ X \to -\Im\ X$, respectively. Therefore, the boundary of $\mathcal{Z}=\mathcal{Y}$ is obtained from Eqs.~\eqref{eq:re-X=Y} and \eqref{eq:im-X=Y} as
\begin{align}
\Re\ X&=1-\frac{2\cos\theta-2\beta}{1-\beta^{2}}\cos\theta, \\
\Im\ X&=-\frac{2\cos\theta-2\beta}{1-\beta^{2}}\sin \theta.
\end{align}

Next, we consider the boundary between the interpretable and non-interpretable regions in the $X$ plane. The boundary with $\mathcal{X}=0$ is given by Eq.~\eqref{eq:X=0} which can be written as
\begin{align}
|1-X|&=1-\gamma|X|, 
\end{align}
with
\begin{align}
\gamma&=\frac{\alpha-1}{\alpha}.
\end{align}
Following a similar calculation with Eq.~\eqref{eq:X=Y-|X|} we obtain:
\begin{align}
|X|&=\frac{2\cos\theta-2\gamma}{1-\gamma^{2}} ,
\end{align}
with $\cos\theta>\gamma$. Equivalently, we have 
\begin{align}
\Re\ X&=\frac{2\cos\theta-2\gamma}{1-\gamma^{2}}\cos \theta, \label{eq:X=0-re}\\
\Im\ X&=\frac{2\cos\theta-2\gamma}{1-\gamma^{2}}\sin \theta. \label{eq:X=0-im}
\end{align}
As discussed above, the boundary $\mathcal{Z}=0$ is calculated with the replacement $X\to 1 - X$:
\begin{align}
\Re\ X&=1-\frac{2\cos\theta-2\gamma}{1-\gamma^{2}}\cos \theta, \label{eq:Z=0-re}\\
\Im\ X&=-\frac{2\cos\theta-2\gamma}{1-\gamma^{2}}\sin \theta. \label{eq:Z=0-im}
\end{align}

Now we consider the classification of the complex $X$ plane with Table~\ref{tab:sum} by using the boundaries obtained above. To determine the dominant component for a given $X$, not all the boundaries are relevant. For instance, the boundary $\mathcal{X}=\mathcal{Y}$ in the $Z$ dominant region is irrelevant, because the boundary classifies the second largest component. In a similar way, the boundaries in the non-interpretable region are also irrelevant. To determine the edge of the boundaries, we calculate $X$, which gives $\mathcal{X}=\mathcal{Y}=\mathcal{Z}$. By substituting $|X| = |1-X|$ into \eqref{eq:X=Y-triangle}, we find the condition $\mathcal{X}=\mathcal{Y}=\mathcal{Z}$:
\begin{align}
|X|&=\frac{\alpha+1}{3} .\label{eq:X=Y=Z}
\end{align}
Because this point also satisfies Eq.~\eqref{eq:reX1/2}, we obtain
\begin{align}
 (\Re\ X,\Im\ X)&=\left(\frac{1}{2},\pm\sqrt{\frac{(\alpha+1)^{2}}{9}-\frac{1}{4}}\right).\label{eq:eq:X=Y=Z-coordinate}
\end{align}
Using $\Re\ X = |X|\cos\theta$, we obtain the corresponding value of the parameter $\theta$ as
\begin{align}
\theta &= \arccos\left[\frac{3}{2(1+\alpha)}\right].
\end{align}
The results obtained here are reflected in Fig.~\ref{fig:alphas}. For example, the imaginary coordinate of the point where the three boundaries intersect increases with $\alpha$. This behavior is readily understood from the analytical expression given in Eq.~\eqref{eq:eq:X=Y=Z-coordinate}.

\section{Determination of $\alpha_{0}$}
\label{sec:alpha-calc}

Here we explain the detailed calculation of $\alpha_{0}$ in Eq.~\eqref{eq:alpha-0}. For the determination of $\alpha_{0}$, we use the ERE where the compositeness of resonances is pure imaginary ($\Re X=0$) as shown in Eq.~\eqref{eq:X-ERE-theta}. From Fig.~\ref{fig:alphas}, with $\alpha>1$, the state with small $\Im X$ is interpretable, while the state becomes non-interpretable for large $\Im X$. In this case, the boundary of the interpretable region is determined by the condition $\mathcal{X} = 0$ in Eq.~\eqref{eq:X=0}, which depends on $\alpha$. The value of $\alpha_{0}$ is determined by requiring that only narrow-width resonances are interpretable and broad resonances are classified as non-interpretable. 

We adopt the criterion~\eqref{eq:interpretable} where the states with $\Re E>\Gamma$ are interpretable. Because the ratio $\Gamma/\Re E$ is determined by the argument of the eigenenergy $\theta_{E}$, the boundary of the interpretable region in the complex $X$ plane is determined by Eq.~\eqref{eq:X-ERE-theta} with $\Im\, E/\Re\, E = (-\Gamma/2)/\Re\, E = -1/2$. In this case, from the relations $\sin\theta_{E} = -1/\sqrt{5}$, $\cos\theta_{E} = 2/\sqrt{5}$, and $\tan\theta_{E} = -1/2$, we find that 
\begin{align}
X&=-i\tan \frac{\theta_{E}}{2}
=i\sqrt{\frac{1-\cos\theta_{E}}{1+\cos\theta_{E}}}
=i\frac{1-\cos\theta_{E}}{-\sin\theta_{E}}
=i(-2+ \sqrt{5}).
\quad \left(\frac{\Im\, E}{\Re\,E} = -\frac{1}{2}\right)
\label{eq:X-}
\end{align}
From Eq.~\eqref{eq:X=0}, the relation between $\alpha$ and $X$ on the boundary is 
\begin{align}
\alpha(X) = \frac{|X|}{|X|-|1-X|+1} . 
\label{eq:alpha-X}
\end{align}
By substituting Eq.~\eqref{eq:X-} into Eq.~\eqref{eq:alpha-X}, $\alpha_{0}$ is obtained as:
\begin{align}
\alpha_{0}&=\frac{-2+\sqrt{5}}{\sqrt{5}-1-\sqrt{10-4\sqrt{5}}}=\frac{\sqrt{5}-1+\sqrt{10-4\sqrt{5}}}{2}.
\end{align}

\section{Trajectory of complex compositeness $X$}
\label{sec:internal}

In this section, we consider the trajectory of the compositeness $X$ in the complex $X$ plane by varying the inverse scattering length $1/a_{0}$ with a fixed negative effective range $r_{e}$. This analysis is performed to understand the behavior of the elementarity $\mathcal{Z}$ in Fig.~\ref{fig:Z-ratio} (a). As shown in Eq.~\eqref{eq:k-pm}, the eigenmomentum $k^{-}$ in the ERE is determined by the scattering length $a_{0}$ and effective range $r_{e}$. At the same time, the compositeness $X$ is also calculated by $a_{0}$ and $r_{e}$ as seen in Eq.~\eqref{eq:wbr}. From these features of the ERE, we can relate the compositeness $X$ to the pole position $k^{-}$. Here, we vary the inverse scattering length from $1/a_{0}\to\infty$ to $-\infty$ with the fixed $r_{e}$ being negative. With this variation of $a_{0}$, the pole $k^{-}$ moves from the bound state to the resonance through the virtual state~\cite{Hyodo:2013iga}. 

\begin{figure}[t]
\centering
\includegraphics[width=0.45\textwidth]{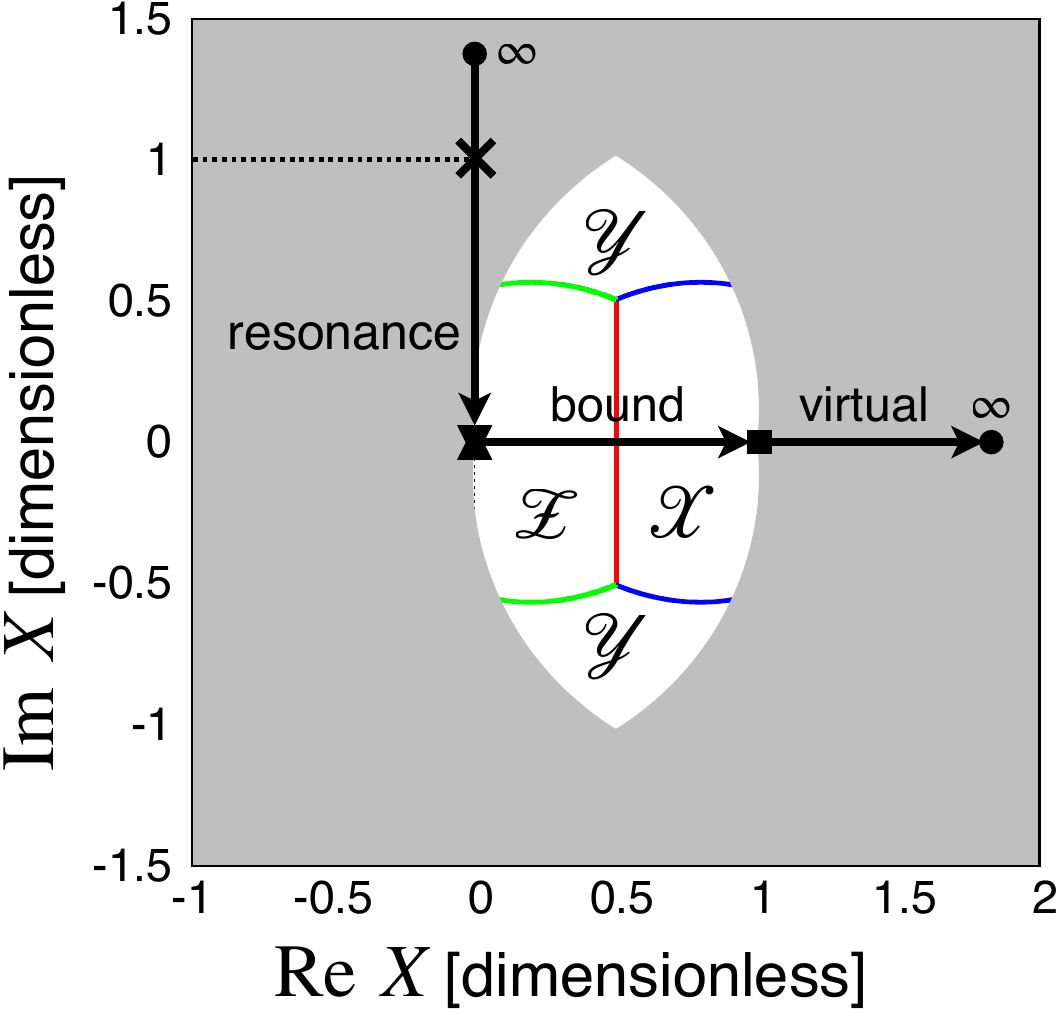}
\caption{The plot of the region in the complex plane where the state is $\mathcal{X,Y}$, or $\mathcal{Z}$ dominant, and the trajectory of the compositeness $X$ with the variation of the pole position (solid line). The shaded region stands for the non-interpretable region with $\alpha = \alpha_{0}$. The triangle, square, circle, cross, and inverted triangle represent the compositeness of states whose eigenmomenta are $k=+\infty i$, $k=0$, $k=i/r_{e}$, $k=(-1+i)/r_{e}$, and $k=+\infty$, respectively.}
\label{fig:trajectory}
\end{figure}

We show the trajectory of the compositeness $X$ by the thick solid line in the complex $X$ plane in Fig.~\ref{fig:trajectory}. The $\mathcal{X,Y,Z}$ dominant regions in the complex $X$ plane are also shown with $\alpha=\alpha_{0}$. The triangle, square, circle, cross, and inverted triangle stand for the values of $X$ whose eigenmomenta are $k^{-}=+i\infty $ ($1/a_{0}=+\infty$), $k^{-}=0$ ($1/a_{0}=0$), $k^{-}=i/r_{e}$ [$1/a_{0}=1/(2r_{e})$], $k^{-}=(-1+i)/r_{e}$ ($1/a_{0}=1/r_{e}$), and $k^{-}=+\infty+i/r_{e}$ ($1/a_{0}=-\infty$), respectively.

At first, we focus on the variation of the compositeness $X$ from the deep bound state with $k^{-}=+i\infty $ (triangle) to the bound state at the threshold with $k^{-}=0$ (square). Because the compositeness of the bound states satisfies $0\leq X\leq 1$, the probability $\mathcal{X}$ is identical with $X$ by definition. The trajectory of the compositeness shows that the internal structure of the bound state is completely elementary dominant $X=\mathcal{X}=0$ for the deep bound state (triangle), which turns from elementary dominant to composite dominant when the pole moves close to the threshold, and finally the bound state becomes completely composite $X=\mathcal{X}=1$ at the threshold (square). This is what the low-energy universality indicates, as shown in Refs.~\cite{Hyodo:2014bda,Hanhart:2014ssa,Kinugawa:2023fbf}.

The virtual states appear on the negative imaginary axis in the complex $k$ plane. When the pole position moves from $k^{-}=0$ (square) to $k^{-}=i/r_{e}$ (circle), the corresponding compositeness $X$ varies in real but $X>1$ region. As shown in Fig.~\ref{fig:trajectory}, the states with $X>1$ are non-interpretable. Therefore, we find that the virtual states do not possess an internal structure that can be interpreted in the present scheme. This is consistent with the fact that the wavefunctions of virtual states have a negative norm~\cite{Braaten:2007nq}; as discussed in Sec.~\ref{sec:eigenstates}, their properties are qualitatively different from those of bound-state wavefunctions.

At $k^{-}=i/r_{e}$, the compositeness $|X|$ diverges~\cite{Hyodo:2013iga}. If the pole position moves closer to $k^{-}=i/r_{e}+i0^{+}$, $X$ goes to $\infty$. On the other hand, $X\to i\infty$ when the pole reaches the divergence point from the complex plane ($k^{-}=i/r_{e}+0^{+}$). Note that when $k^{-}=i/r_{e}$, the scattering length is $1/a_{0}=1/(2r_{e})$, and the other pole also comes to the same position $k^{+}=i/r_{e}$ as seen in Eq.~(2). This means that the two eigenstates with $k^{\pm}$ coalesce, and such a special point in the parameter space is called the exceptional point~\cite{Moiseyev,Bergholtz:2019deh}.

As $1/a_{0}$ is large and negative, the pole moves to the fourth quadrant with non-zero Re$\,~k^{-}$. When the real part of the eigenmomentum is larger than the magnitude of the imaginary part (Re $k^{-}>$ |Im $k^{-}|$), the pole becomes a resonance. When $0 < \Re\, k^{-} < |\Im\, k^{-}|$, the compositeness $X$ is calculated as purely imaginary in the ERE as discussed around Eq.~\eqref{eq:X-ERE-theta}. As the real part of $k^{-}$ increases, the argument $|\theta_{k}|$ decreases, and |Im $X$| decreases from $\infty$ and reaches unity at $\Re\, k = |\Im\, k|$ (cross). In this case, the probability $\mathcal{X}$ is smaller than zero, and the eigenstates are regarded as non-interpretable in this study. 

Finally, we discuss the resonance poles in the Re$\,k^{-}>$ |Im$\,k^-|$ region. The corresponding trajectory of $X$ starts from $X = i$ (cross), and ends at $X = 0$ (inverted triangle) as the ratio $\Gamma/\Re\, E$ decreases. When $X = i$, $\Re\, E = 0$ with the diverging ratio $\Gamma/\Re\, E \to \infty$ and hence such a state is classified as non-interpretable, as shown in Fig.~\ref{fig:trajectory}. On the other hand, when the ratio decreases sufficiently ($\Gamma/\Re\, E \leq 1$), the resonance enters the interpretable and elementary-dominant region. In the $\Gamma/\Re\, E \to 0$ limit, the imaginary part of the pole position becomes negligible compared with its real part, and the pole effectively approaches the real axis. The state can therefore be regarded as an isolated state. The compositeness is zero in this case, and the isolated state is interpreted as completely elementary. This is consistent with the expectation that narrow resonances have a large negative effective range.

\end{document}